\documentclass{aa}  
\usepackage{natbib}
\usepackage{ragged2e}
\usepackage{mathtools}
\usepackage{txfonts}

\usepackage{ulem}
\usepackage{newtxtext,newtxmath}
\usepackage[T1]{fontenc}
\usepackage{ulem}
\usepackage{xcolor} 
\usepackage{graphicx}	% Including figure files
\usepackage{amsmath}	% Advanced maths commands
\usepackage{amssymb}	% Extra maths symbols
\usepackage{float}
\usepackage{multicol}  
\usepackage{multirow}        % Multi-column entries in tables
\usepackage{bm}		% Bold maths symbols, including upright Greek
\usepackage{pdflscape}	% Landscape pages
\usepackage{rotating}
\usepackage[colorlinks=true,citecolor=blue, urlcolor=blue, filecolor=blue, linkcolor=blue]{hyperref}%
\usepackage{enumerate}
\usepackage{fontawesome}
\defcitealias{1989ApJ...344..115C}{CH89}
\def\mbh{$M_{\rm BH}$\/}

\def\lledd{$L_\mathrm{bol}/L_\mathrm{Edd}$}

\def\civ{{\sc{Civ}}$\lambda$1549\/}
\def\civonly{{\sc{Civ}}\/}
\def\ciii{{\sc{Ciii]}}$\lambda$1909\/}

\def\aliii{{Al\sc{iii}}$\lambda$1857\/}
\def\rfeopt{$R_{\rm FeII,opt}$\/}

\def\feiiq{\rm Fe{\sc ii}$\lambda$4570\/}
\def\msol{M$_\odot$\/}

\def\ltsima{$\; \buildrel < \over \sim \;$}
\def\ltsim{\lower.5ex\hbox{\ltsima}}  
\def\gtsima{$\; \buildrel > \over \sim \;$}

\def\gtsim{\lower.5ex\hbox{\gtsima}} 

\def\lya{{ Ly}$\alpha$}
\def\oiii{[\ion{O}{iii}]$\lambda\lambda$4959,5007\/}
\def\oiiiseven{[\ion{O}{III}]$\lambda$5007\/}

\def\ha{{\sc{H}}$\alpha$\/}
\def\hb{{\sc{H}}$\beta$\/}

\def\mgii{{Mg\sc{ii}}$\lambda$2800\/}

\def\mgiionly{{Mg\sc{ii}}\/}

\def\oiiiopt{{\sc{[Oiii]}}$\lambda\lambda$4959,5007\/}

\def\oi{{\sc{[Oi]}}$\lambda\lambda$6302,6365\/}

\def\nii{{\sc{[NII]}}$\lambda\lambda$6549,6585\/}

\def\sii{{\sc{[SII]}}$\lambda\lambda$6718,6732\/}

\def\siiii{{Si\sc{iii}]}$\lambda$1892\/}

\def\feiiuv{{Fe\sc{ii}}$_{\rm UV}$\/}
\def\feiiopt{{Fe\sc{ii}}$_{\rm opt}$\/}
\def\feii{{Fe\sc{ii}}\/}
\def\fe{{\sc{Fe}}\/}

\def\fe76087{{\sc [Fe vii]}$\lambda$6087\/}
\def\kms{km\,s$^{-1}$}

\def\rk{$R_{\rm K}$\/}

\def\heii{{{\sc H}e{\sc ii}}$\lambda$4686\/}

\def\o4959{{\sc{[Oiii]}}$\lambda$4959\/}
\defcitealias{2023MNRAS.525.4474M}{STM23}
\definecolor{darkorange}{rgb}{1,0.612,0}
\definecolor{aquamarine}{rgb}{0.498,1,0.8314}
\begin{document} 
	\authorrunning{}
	
	\title{The Quasar \object{3C 47}: Extreme Population B Jetted Source With Double-peaked Profile\thanks{Based on observations collected at the Centro Astronómico Hispano en Andalucía (CAHA) at Calar Alto, operated jointly  by the Junta de Andalucía and Consejo Superior de Investigaciones Científicas (IAA-CSIC).}}
	
	\author{Shimeles Terefe Mengistue\inst{1,2,3}\thanks{Visiting researcher at the IAA-CSIC, Spain as a PhD fellow.}
		\and
		Paola  Marziani\inst{4}
		\and 
		Ascensi\'{o}n del Olmo\inst{5}
		\and
		Mirjana  Povi\'{c}\inst{1,5,6}
		\and
		Jaime  Perea\inst{5}
		\and
		Alice  Deconto Machado\inst{5}   
	}

	\institute{Space Science and Geospatial Institute (SSGI),  Entoto Observatory and
		Research Centre (EORC), Astronomy and Astrophysics Department, 
		P.O.Box 33679, Addis Ababa, Ethiopia.\\
		\email{shimeles11@gmail.com} 
		\and
		Addis Ababa University (AAU), P.O.Box 1176, Addis Ababa, Ethiopia. 
		\and
		Jimma University, College of Natural Sciences, Department of Physics, P.O.Box 378, Jimma, Ethiopia. 
		\and    
		Istituto Nazionale di Astrofisica (INAF), Osservatorio Astronomico di Padova,vicolo dell$^{’}$ Osservatorio 5, Padova I-35122, Italy.
		\and
		Instituto de Astrofisica de Andaluc\'{i}a (IAA-CSIC), Glorieta de la Astronomia s/n, Granada E-18008, Spain.
		\and
		Mbarara University of Science and Technology (MUST), Faculty of Science, Physics Department, P.O. Box 1410, Mbarara, Uganda.              
	}
	
	\date{Received XX, 2023; accepted XX, 2023}
	
		%\abstract {} {Text of aims} {Text of methods} {Text of results} {} and 300 words
	\abstract
{An optically thick, geometrically thin accretion disk (AD) around a supermassive black hole might contribute to broad-line emission in type-1 active galactic nuclei (AGN). However, emission line profiles are most often not immediately consistent with the profiles expected from a rotating disk. The extent to which an AD in AGN contributes to the broad Balmer lines and high-ionization UV lines in radio-loud (RL) AGN needs to be investigated.	}
{ This work aims to address whether the AD can account for the double-peaked profiles observed in the Balmer lines (\hb, \ha), near-UV (\mgii), and  high-ionization UV lines (\civ, \ciii)  of the extremely jetted quasar \object{3C 47}.}
{The low ionization lines (LILs) (\hb, \ha, and \mgii) were analyzed using a relativistic Keplerian AD model. Fits were carried out following  Bayesian and multicomponent non-linear approaches. The profiles of prototypical high ionization lines (HILs) were also modeled by the contribution of the AD, along with fairly symmetric additional components.}
{ The LIL profiles of \object{3C 47} are in very good agreement with a relativistic Keplerian AD model. The disk emission is constrained between $\sim 10^2$ and $\sim 10^3$\ gravitational radii, with a viewing angle of $\approx$ 30 degrees. }
{The study provides convincing direct observational evidence for the presence of an AD and explains the HIL profiles are due to disk and failed wind contributions. The agreement between the observed profiles of the LILs and the model is remarkable. The main alternative, a double broad line region associated with a binary black hole, is found to be less appealing than the disk model for the quasar \object{3C 47}.
	}
	
	
\titlerunning{\object{3C 47}, Jetted source with double-peaked profile}
	\authorrunning{S. T. Mengistue et al.}
	
	\maketitle
%-------------------------------------------------------------------
	\section{Introduction}
	\label{sec:intr}
	%እግዚአብሔር ይመስገን
 
In the current working picture of active galactic nuclei (AGN), the underlying power source is thought to be a supermassive black hole (SMBH). An accretion disk (hereafter AD) around the central SMBH is expected in all the AGN. ADs provide an efficient mechanism for dissipating the angular momentum of the accreting matter via viscous stresses \citep[e.g.,][] { 1973A&A....24..337S, 2021SSRv..217...12D, 2023arXiv230207925L}. In addition, they are regarded as an essential ingredient for the production of relativistic jets, which are observed in a considerable fraction of AGN \citep[e.g.,][]{1990agn..conf..161B,2019ApJ...877..130S,2019ARA&A..57..467B,2021AN....342..142C, 2022Univ....8...85M}. 
High-density material in an AD may be required for the production of the low-ionization lines (LILs) \citep[][]{1997ASPC..113...56R,2019MNRAS.490.1738Z,2020ApJ...903...31H}. Emission from the surface of a photoionized, relativistic, Keplerian AD  produces profiles of double-peaked lines with two distinctive features: (1) a stronger blueshifted peak due to Doppler boosting; (2) a redward shift - increasing toward the line base - associated with gravitational redshift and Doppler transverse effect  \citep[][hereafter \citetalias{1989ApJ...344..115C}]{1989ApJ...344..115C}.  
In this respect, the most commonly proposed solution for double-peaked emission from the AD is the assumption of an elevated structure around the inner disk that would illuminate the outer disk and drive the line emission \citep{2003AJ....126.1720S,2003ApJ...599..886E,2019MNRAS.486.1138R}.\\
\begin{table*}[htp!]
\centering
\caption{ Summary of general optical and radio properties. }
\label{tab:tabr}
\setlength{\tabcolsep}{3.9pt}
\scalebox{0.78}{
\begin{tabular}{l cc ccccccc c ccccc} 
\hline\hline 
\rule{0pt}{2.5ex}% inserting double-line
& & & \multicolumn{7}{c}{ Optical}& & \multicolumn{5}{c}{ Radio}\\[0.5ex] \cline{4-10} \cline{12-16}
\rule{0pt}{2.5ex}
Object &\multicolumn{2}{c}{ Coordinates} & \multicolumn{1}{c}{z}& $ V $ & $ M_{B}$ & \multicolumn{1}{c}{Date of} & \multicolumn{1}{c}{Total exp.} & Air mass & S/N & & NVSS  & $\alpha$ & log\rk & log$(P_{\nu})$ & Morphology\\
& RA(2000) & Dec(2000) &  &  &  & observation & time (s) & & & & Flux (mJy) & & & (ergs\, s$^{-1}$\,Hz$^{-1}$) & \\
\vspace{0.20cm}
(1) & (2)& (3) & (4) & (5) & (6) & (7) & (8) & (9) &(10) & & (11) & (12) & (13) & (14) & (15)\\ 
\hline 
\\ 
\object{3C 47} & 01 36 24.41 & +20 57 27.44 & 0.4248 & 18.1 & -23.3 & 22-10-2012 & 3600 & 1.067 & $\approx 60$ & & 3900.7 & 0.9 &  $\approx 4$ & 34.5 & LD\\[1ex]
\hline  \\
\end{tabular} 	}

{\small{\raggedright\textbf{Notes}: { Col. 2 is in units of hours, minutes, and seconds. Col. 3 is in units of degrees, minutes, and seconds. The radio spectral index $\alpha$ in Col. (12) is  from \citet{2010A&A...511A..53V}.  LD: lobe dominate, with core-to-lobe ratio $\approx 0.11$\ at 1.4 GHz \citep{1994AJ....108..766B}.} \par}}
\end{table*}
	
\indent Even if the majority of AGN spectra are single-peaked, there are pieces of evidence about a double-peaked structure that can be acquired from observations of very broad, double-peaked emission lines and identification of asymmetries and substructure in the line profiles \citep[e.g.,][]{2002A&A...390..473P,2003A&A...407..461K,2004mas..conf..199S}. \citet{1979ApJ...228L..55M}  found that powerful radio galaxies and radio-loud (RL) quasars with extended radio morphology have the broadest and most complex Balmer line profiles and are preferred hosts of double-peaked emitters.  
AGN with double-peaked emission lines are an interesting class of objects,  though only a small fraction has been found \citep[e.g.,][]{1994ApJS...90....1E, 2003ApJ...599..886E,2003AJ....126.1720S,2009NewAR..53..133E,2023MNRAS.524.5827F}. According to \citet{2005ApJ...625L..35W}, these double-peaked lines are among the broadest optical emission lines, with a full width at half maximum (FWHM) in some cases exceeding 15000 \kms. Following the classification scheme of \citet{2002ApJ...566L..71S}, they are more frequently found in extreme Population B, and, more precisely, in spectral type (ST) B1$^{++}$. This ST is populated by a tiny minority of quasars in optically selected samples \citep{2013ApJ...764..150M}, and consistently associated with FWHM \hb\ in the range
 of 12000 -- 16000 \kms, and low or undetectable singly-ionized (\feii) emission. They are evolved sources with spectacular extended narrow line regions (NLRs) and a high prevalence of powerful radio jets \citep{2008MNRAS.387..856Z,2019A&A...630A.110G}. In physical terms, sources belonging to this ST are seen at a relatively high inclination and/or have large SMBH mass (\mbh), and very low Eddington ratio (\lledd) ($\sim 0.01$) \citep[e.g., ][]{2019ApJ...882...79P}.

A variety of mechanisms have been proposed to explain the origin of double-peaked emission line profiles and their unique kinematic signature, beyond relativistic motions in an AD \citep[e.g., \citetalias{1989ApJ...344..115C};][]{ 1990ApJ...354L...1C},  such as: two separate broad-line regions (BLR) as a signature of binary black holes \citep{1983LIACo..24..473G}, a biconical outflow \citep{1990ApJ...365..115Z}, or a highly anisotropic distribution of emission line gas \citep{1995ApJ...453L..87W, 1996ApJ...469..113G}.
A careful consideration of the basic physical arguments and recent observational results, tend to confirm that the most likely origin of double-peaked emission lines is the AD, and on the converse, these double-peaked profiles provide dynamical evidence about the structure of the AD \citep[e.g.,][]{1997ApJ...483..194C,1998AdSpR..21...33E, 2000ApJ...541..120H, 2003AJ....126.1720S, 2003ApJ...599..886E,2011NewA...16..122B,2011MNRAS.416.2857Z,2017MNRAS.472L..99L,2019MNRAS.486.1138R,2019ApJ...877...33Z}.  Several additional observational tests combined with physical considerations also favored the AD origin over other possibilities \citep[e.g.,][]{1994ApJS...90....1E,1998AGAb...14...61D,1999ASPC..175..163E,2003ApJ...599..886E,2003AJ....126.1720S,2009NewAR..53..133E,2013MNRAS.431L.112Z,2021PASJ...73..596W,2017ApJ...835..236S}. 
	
Nonetheless, open questions concerning the disk emission in this class of rare AGN remain.  UV spectroscopy of double-peaked emitters with the Hubble Space Telescope (HST) did not provide unambiguous evidence for an AD origin \citep[e.g.,][]{1996ApJ...464..704H,1998AdSpR..21...33E,2004IAUS..222...29E, 2011MNRAS.416.2857Z}, and high-ionization lines (HILs) (e.g., Ly${\alpha}$ and \civ) frequently lack double-peaked profiles \citep[e.g.,][]{1996ApJ...464..704H,1997ApJ...474...91M,2004IAUS..222...29E,2009NewAR..53..133E}.

 \begin{figure*}[tp!]
		\centering
		\includegraphics[width=\textwidth]{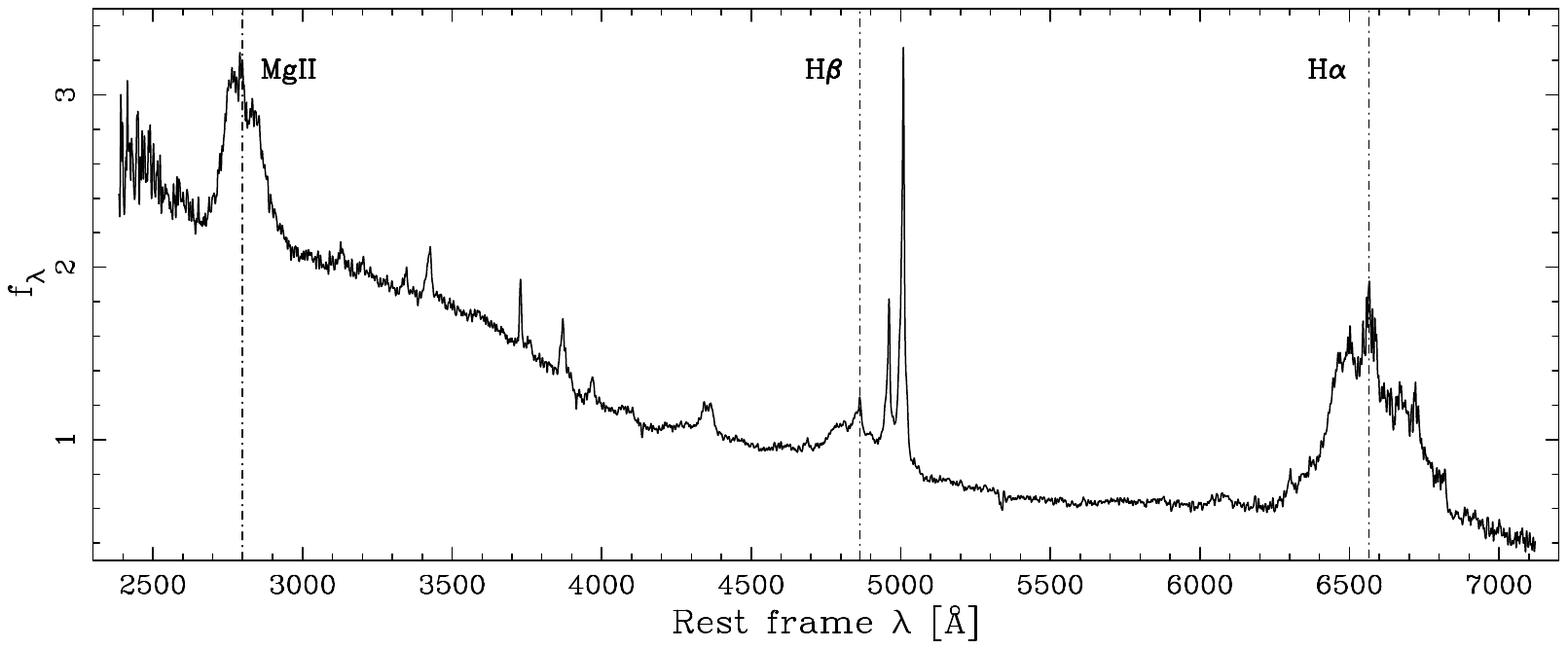}	
\vspace{-5.3cm}
\caption{ A new rest-frame spectrum of \object{3C 47} that covers \mgiionly, \hb\ and \ha. Abscissa is wavelength in \AA\ and the ordinate corresponds to specific flux in units of 10$^{-15}$ergs\ s$^{-1}$cm$^{-2}$\AA$^{-1}$. 
  The dot-dashed vertical lines trace the rest-frame vacuum wavelength.} 
 \label {fig:restf}
\end{figure*}

To evaluate the extent to which AD is a source of the broad Balmer lines and UV HILs in RL AGN, we focused on a strong RL quasar, \object{3C 47} \citep[e.g., ][]{2010A&A...523A...9H}.
Previous studies on \object{3C 47} indicated that it exhibits peculiar broad emission line profiles with multiple components, making it a member of the AGN class of double-peaked emitters \citep{2003ApJ...599..886E}. 
This work presents new simultaneous optical and near-UV spectra of the relativistically jetted double-peaked quasar, \object{3C 47} at redshift 0.4248 from long-slit spectroscopic observations. These new observations of \object{3C 47} yielded a spectrum with a high signal-to-noise (S/N) ratio, high resolution, and broad and strong blue and red peaks that are typical indicators of double-peaked emitters in the Balmer lines (\hb\ and \ha) as well as the near-UV \mgii\ (hereafter \mgiionly) line. In addition, we also present and interpret the semi-forbidden lines of the $\lambda 1900$ \AA\ blend dominated by \ciii\ and the HIL \civ\ from the HST Faint Object Spectrograph (HST-FOS) archive.
 
 The paper is organized as follows: the data used in this work are described in Sect. \ref{sec:data}. Spectral non-linear multicomponent and model fittings, as well as full broad profile analysis, are described in Sect. \ref{sec:SA}. The AD model fit, following the approximation by \citetalias{1989ApJ...344..115C} was carried out by using Bayesian methods for each of the three spectral regions,  \mgiionly, \hb\ and \ha, as described in Sect. \ref{sec:SA}. The results of using a model that -- for the first time in this object -- attributes the emission of the double-peaked lines to the surface of a photoionized, relativistic Keplerian AD plus the results of the broad profile parameters, are described in Sects. \ref{sec:SA} and \ref{sec:results}. We discuss the resulting accretion parameters, the interpretation of the \civ\ and \ciii\ emission lines, the explanation of the \civ\ profile as due to the contribution of the AD and a failed wind, and the available alternative models in Sect. \ref{sec:disc}.   Finally,  our conclusions are given in Sect. \ref{sec:concl}. Throughout this  paper,  we adopt a flat $\Lambda$CDM cosmology with $\Omega_{\Lambda}$= 0.7, $\Omega_{0}$= 0.3, and
$H_{0} = 70\,$km\,s$^{-1}$Mpc$^{-1}$.

\section{Data}
\label{sec:data}
\subsection{{Near-UV and optical observations}}
\label{sec:obser}

New long-slit optical spectroscopic observations were obtained on the night of 22 October 2012 with the TWIN spectrograph of the 3.5m Cassegrain telescope at the Calar Alto Observatory (CAHA, Almería, Spain)\footnote{\url{http://www.caha.es/}}. Among the sample of 12 extremely jetted RL quasars in the observations of \citet[][hereafter \citetalias{2023MNRAS.525.4474M}]{2023MNRAS.525.4474M}, \object{3C 47} is the one found to possess a double-peaked broad line profile and was therefore singled out for the more detailed analysis presented in this paper. The observations were obtained with a slit width of 1.2 arcsec. The total exposure time was split into three exposures (each of 1200 seconds) to eliminate cosmic rays by combining different exposures. The mean air mass was found to be 1.09. 
We obtained high quality spectra with S/N of $\approx$ 60 in the continuum near \hb\ and \ha\ lines. The observational setup and data reduction of \object{3C 47} are the same as the ones of the other 11 sources described in \citetalias{2023MNRAS.525.4474M}. The spectra of  \object{3C 47} reveal three spectacular emission lines corresponding to \hb\ and \ha{} in the red and \mgiionly\ in the blue arm of the spectrograph.
	
We determined the median spectroscopic redshift of  \object{3C 47} by the method detailed in \citetalias{2023MNRAS.525.4474M} (Sect. 2.2) in which we measured the observed vacuum wavelength of individual narrow emission lines (e.g., [OII]$\lambda$3727, H$\gamma$, \hb\ and \oiiiopt\/) available in the spectra and found it to be $z \approx$ 0.4248 $\pm$ 0.0004 in fair agreement with the previously measured value of $z$\,$\approx$\, 0.4255 \citep{2010A&A...518A..10V}. Basic observational and physical properties of this source are summarized in Table \ref{tab:tabr} from Cols. 2 - 10. The rest-frame spectra that provide a concomitant coverage of \mgiionly, \hb\, and \ha\ regions are shown in Figure \ref{fig:restf}. \object{3C 47} has an apparent Johnson magnitude $V \approx$ 18.1 and is hence relatively bright. The absolute magnitude  ($M_{B}$) of -23.3 \citep{2010A&A...518A..10V} makes it a rather low luminosity AGN.
	
	\subsection{Archival data}
	\label{sec:arch}
	
	\subsubsection{Radio  data}
	
	\object{3C 47} is a classical, powerful Fanaroff-Riley II radio source \citep{1974MNRAS.167P..31F}, known for a spectacular double-lobed structure with a strong core \citep{1984ApJ...283..515B}. 
	Some basic radio properties, including the radio flux density obtained at 1.4\,GHz (21\,cm) from the National Radio Astronomy Observatory (NRAO) Very Large Array (VLA) Sky Survey (NVSS)\footnote{NVSS-\url{https://www.cv.nrao.edu/nvss/NVSSPoint.shtml}}  \citep{1998AJ....115.1693C} and the radio spectral index $\alpha$ from \citet{2010A&A...511A..53V}  are reported in Table \ref{tab:tabr}  (Cols. 11 and 12, respectively). The value listed for the specific flux is the sum of two sources detected by the NVSS, corresponding to the two major lobes. The radio loudness parameter, which is the ratio of the rest-frame radio flux density at 1.4 GHz to the rest-frame optical flux density in the g-band, \rk{} \citep[e.g.,][]{2008MNRAS.387..856Z, 2015MNRAS.452.3776G} as well as the radio power ($P_{\nu}$) using the relation from \citet[][their equation 1, also detailed in \citetalias{2023MNRAS.525.4474M}, Sect. 2.3]{2008ApJ...687..859S}  are listed in 
 Cols. 13 and 14. In all the calculations, we used the convention for the spectral index, $S_\nu \propto \nu^{-\alpha}$. \object{3C 47} is among the extreme RL sources with log\rk $\sim$ 4 and high radio-power, log$P_{\nu} \approx$  34.5\,[${\rm ergs\,s^{-1}\,Hz^{-1}}$]. These properties are common among double-peakers. Double-peaked emitters are more likely to belong to RL AGN  and they didn't show special radio properties \citep[e.g.,][]{1998AdSpR..21...33E, 2003AJ....126.1720S}. Further details on the relation between optical and radio properties are provided in Sect. \ref{radio-opt}.

\subsubsection{UV  data}
	
\begin{figure*}[htp!]
		\centering
		  \includegraphics[width=\textwidth]{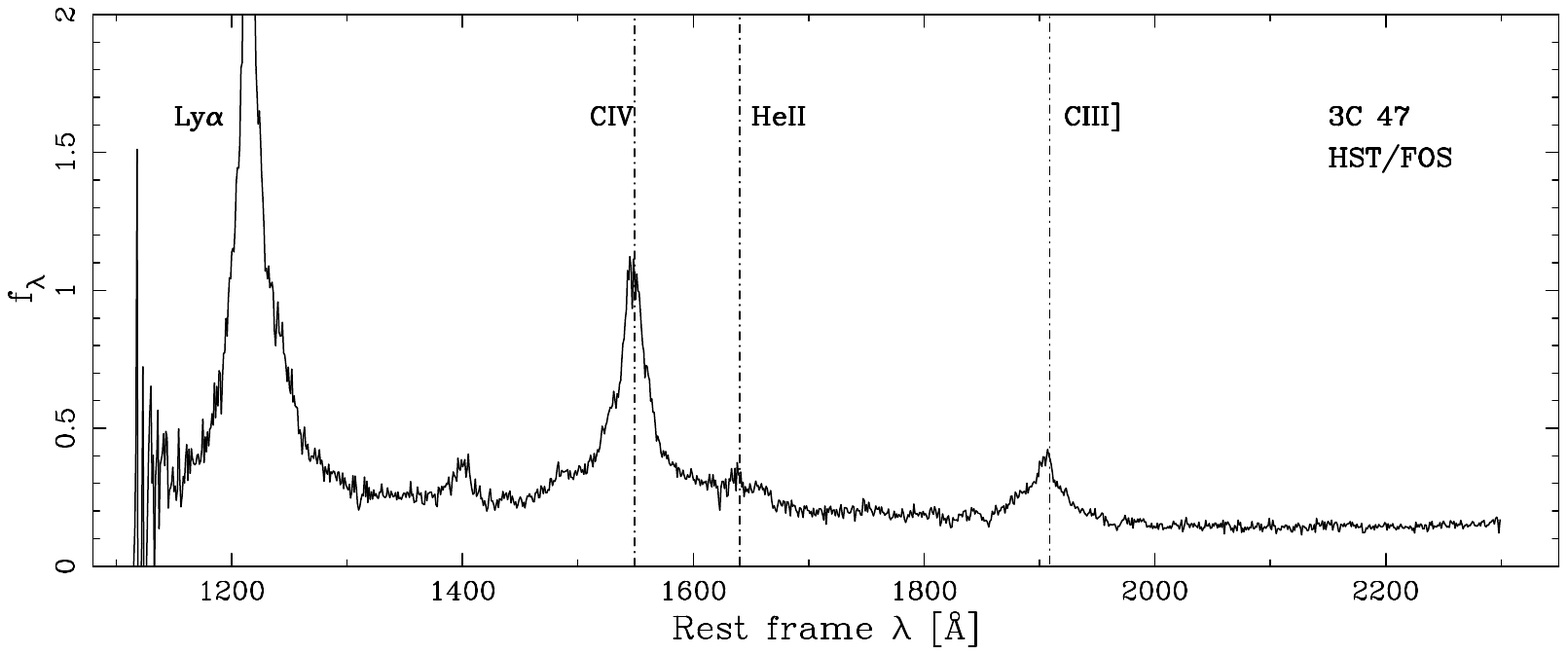}
  \vspace{-5.5cm}
       \caption{ HST/FOS rest-frame spectrum of \object{3C 47} that covers \lya, \civ, HeII$\lambda$1640, and \ciii. The dot-dashed vertical lines trace the rest-frame vacuum wavelength. The ordinate corresponds to specific flux in units of 10$^{-15}$ergs\ s$^{-1}$cm$^{-2}$\AA$^{-1}$.} 
		
		\label {fig:restuv}
	\end{figure*}

  The UV data have been obtained from the HST-FOS \footnote{HST-FOS-\url{https://archive.stsci.edu/missions-and-data/hst}} archive. 
 The resulting rest-frame UV spectra,  covering \civ{}, HeII$\lambda$1640, and \ciii{} regions are shown in Figure \ref{fig:restuv}. The detailed disc profile analysis for this band is presented in Sect. \ref{subsec:model} and discussed in Sect. \ref{sec:failed}.
	
	\subsubsection{Optical data}
	
	 To examine an alternative scenario (detailed in Sect. \ref{sec:alter}) as an explanation of the observed double-peaked profile, two additional previously observed optical spectra  were retrieved,
 namely, an ESO spectrum obtained at the 1.52m telescope on October 14, 1996 \citep{2003ApJS..145..199M}, and a spectrum obtained at the Copernico telescope equipped with AFOSC on Nov. 27, 2006, covering \hb\ and \mgiionly\ \citep{2008agn8.confE..14D}. 
	
	\section{Spectral analysis}
	\label{sec:SA}
 We implemented two types of fittings for each one of the spectral regions: 
i) a model fitting by using a relativistic keplerian AD model, and ii) an empirical 
multicomponent non-linear fitting of the continuum and the emission features by using IRAF task {\tt specfit} \citep[][]{1994ASPC...61..437K}.

\subsection{Accretion disk model}
	\label{subsec:model}	
In this work, we applied the model of \citetalias[][]{1989ApJ...344..115C}, in which AD emission surrounding a single SMBH is assumed to be the origin of double-peaked broad emission lines observed in our spectra. 
The model assumes that a uniform axisymmetric disk model produces double-peaked line profiles, with the blue peak stronger than the red peak due to Doppler boosting.  In the \citetalias[][]{1989ApJ...344..115C} model, the emission is driven by illumination of the outer part of the disk from an elevated,  or vertically extended structure.  This structure could be an ion-supported torus,  that photoionizes the geometrically thin outer disk to produce the observed disk line profiles. This ion torus emits harder than that of the standard, geometrically thin, optically thick AD and such property may overcome difficulties associated with the energy budget of the AD. Other illumination geometries may also be possible \citep[e.g.,][]{1990A&A...229..302D}.	\\

 We used the integral expression for the line profile of an optically thick AD (Eq. 7 of \citetalias{1989ApJ...344..115C}) to fit the double-peaked profiles observed in our spectra.  
The model includes five crucial and freely varying parameters: the two line emitting portions at different radii, the inclination angle, the broadening parameter, and the line emissivity index. The line-emitting portion of the disk is assumed to be an annulus with inner and outer disk radii $\xi_{1}$ and $\xi_{2}$ respectively, in units of gravitational radius, $r_{g}=G M_{BH}/c^{2}$.

This model also assumes that the double-peaked emission line originates from the surface whose axis is inclined by an angle $\theta$ relative to the line of sight, between the two radii. Electron scattering or turbulent motion is assumed to be the cause of line broadening and is represented by a Gaussian profile of velocity dispersion $\sigma$ expressed in \kms\ in the rest-frame of the emitter. An axisymmetric emissivity that varies continuously is represented as a function of the radius of the disk ($\it{Q}$) and a dimensionless power law index $\it{q}$ as $Q=\xi^{-q}$.
	
\subsection{Bayesian fit to the accretion disk model}
\label{sec:bayesian}
	
	The fit of the observed broad line profiles of \object{3C 47} to the AD model,  was carried out by using the Bayesian method, independently for each spectral region: \mgiionly, \hb{} and \ha. 
 
 The input double-peaked broad emission lines were obtained after removing the other components: 
  i.e., mainly the power law of the continuum, the \feii\  template, and the narrow components, as explained in Sec. \ref{sec:empiricalhb}.
	
First estimation of the parameters $\{q,\ \sigma/\nu,\ \xi_1,\ \xi_2,\ \theta \}$  was obtained by evaluating the $\chi^2$ of the fit of the broad profiles to the function defined by Eq. (7) of \citetalias[][]{1989ApJ...344..115C} for several thousands of combinations. 
This was done to obtain a physically possible range of values compatible with the observed double-peaked profiles, as well as initial estimations for each parameter. After inspecting the distribution of $\chi^2$ for the \mgiionly, \hb\, and \ha\ we can ensure that, for values outside the range of the parameters given in Table \ref{tab:modpar}, $\chi^2$ grows monotonically outwards. A non-linear least squares (as well as a maximum likelihood) fit is obtained by using the midpoints of the ranges as initial values, resulting in parameter solutions that are contained well inside the ranges in all the lines.

In addition, we applied a Monte Carlo Markov Chain (MCMC) approach, where we obtained model approximations using the {\tt{emcee}}: The MCMC Hammer package \citep{2013ascl.soft03002F}, which is based on slight modifications to the Metropolis-Hastings method. For the Bayesian inference, we used two different prior functions ($p(\eta)$): 1) a uniform prior for each AD model parameter with the range of values listed in Table \ref{tab:modpar},
	\[
	p(\eta) = 
	\begin{dcases}
		\frac{1}{\eta_2 - \eta_1},& \text{if } \eta_1 \leq \eta < \eta_2 \\
		0,              & \text{otherwise}
	\end{dcases}
	\]
	
	\noindent and 2) Gaussian distribution priors with central values corresponding to the midpoints described previously, and $\sigma$ guaranteeing the full range of parameters to be sampled
	\[
	p(\eta) = \frac{1}{\sigma_{\eta}\sqrt{2\pi}} 
	e^{-\frac{(\eta-\eta_c)^2}{2 \ \sigma_\eta^2}}
	\]
	with $\eta_c$ and $\sigma_{\eta}$ describing the center and spread of each probability $p(\eta)$ prior. 
 In such a way that the total prior is the product of each parameter prior distribution $p(\eta)$: 
	\[
	\prod_{\eta} p(\eta), {\rm with}\  \eta \in \{q,\ \sigma/\nu,\ \xi_1,\ \xi_2,\ \theta \}
	\]
 
	\begin{table}
		\centering
		\caption{Model parameters input list.}
		\label{tab:modpar}
		\scalebox{1.1}{
			\begin{tabular}{@{\extracolsep{4.5pt}}l c}   
				\hline\hline  \rule{0pt}{2.5ex}
				Parameters & Range of values \\
				%  (1) & (2)& (3) \\
				\hline \rule{0pt}{2.5ex}
				\ \ q & 1.3 -- 2.5 \\[1ex]
				\ \ $\sigma/\nu_{0}$ &1.5x10$^{-3}$--1.1x10$^{-2}$ \\[1ex]
				\ \ $\xi_{1}(r_{g})$ & 40 -- 300 \\[1ex]
				\ \ $\xi_{2}(r_{g})$ & 100 -- 2000 \\[1ex]
				\ \ $\theta$ (degrees) & 15 -- 42 \\[0.5ex]
				\hline                             
			\end{tabular} 
		}%\\[0.5ex]
	\end{table}
	
\begin{figure*}
		\centering 
		\includegraphics[width=0.48\textwidth]{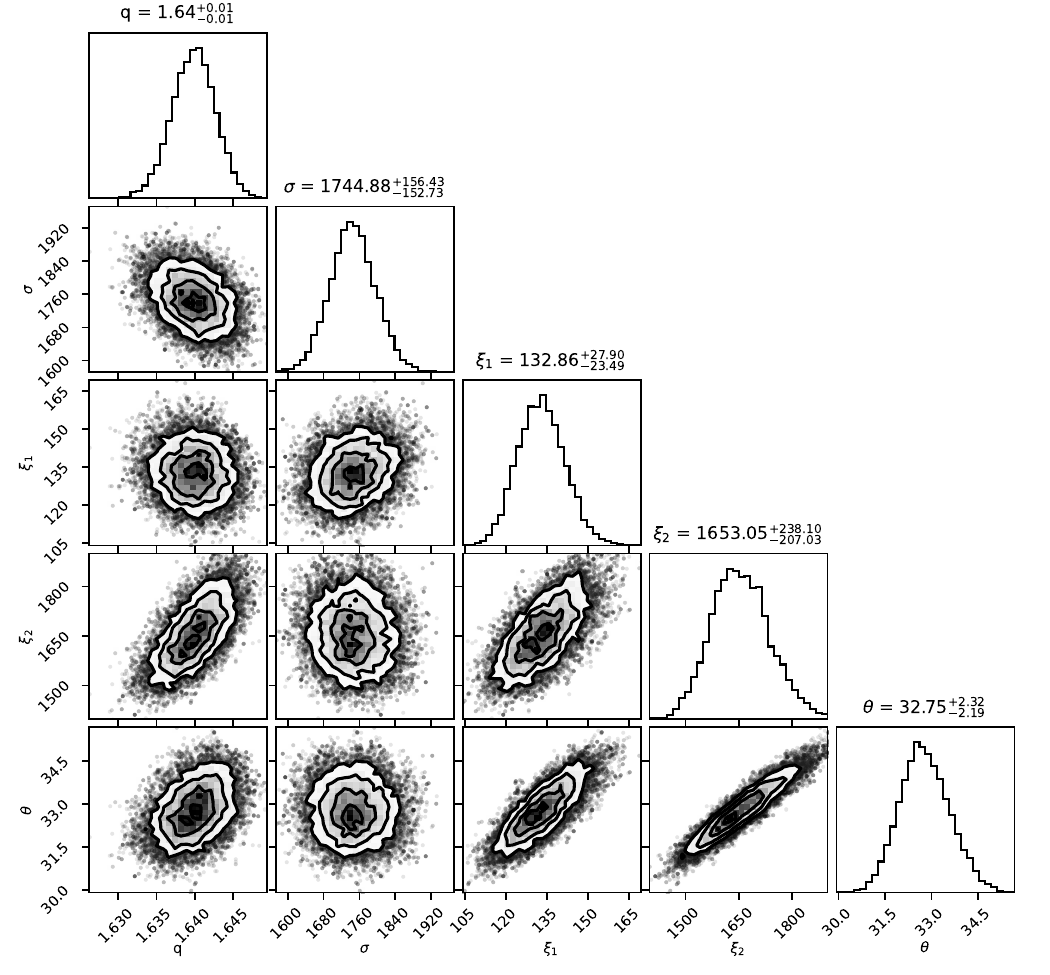}
		\includegraphics[width=0.40\textwidth]{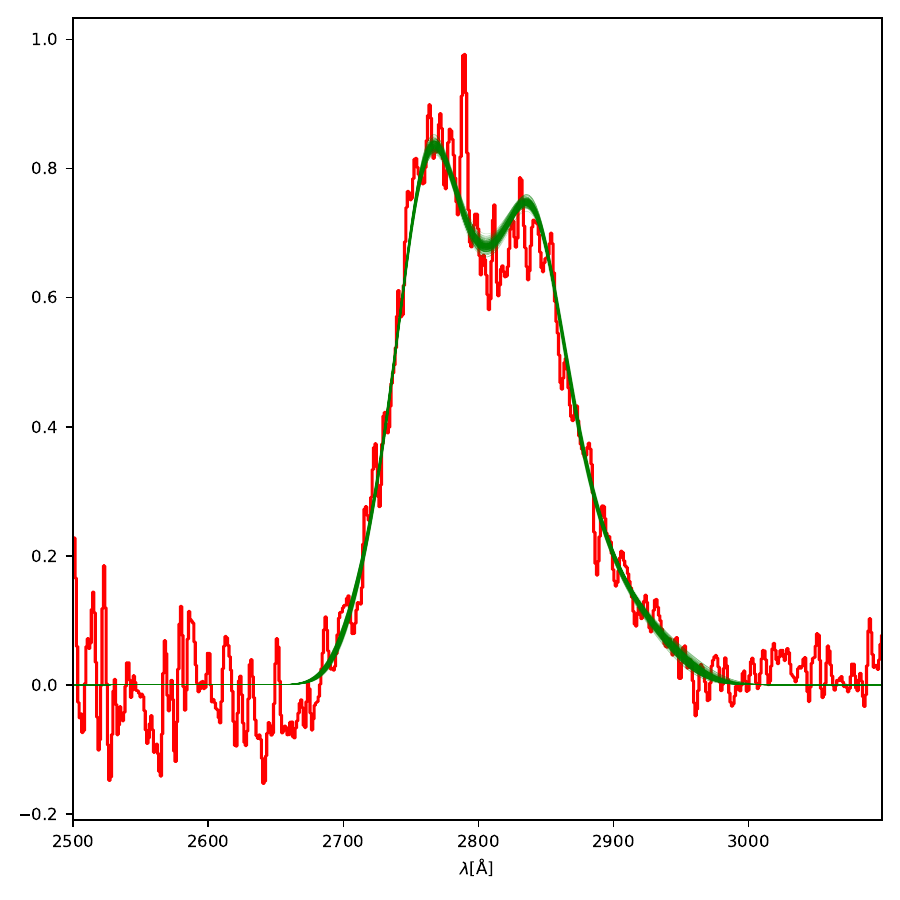}
		
		\caption{Left: Corner plot showing as histograms the posteriors of the five AD parameters for \mgiionly,  and covariance maps between the parameters. Right: 250 randomly selected posterior solutions from the Bayesian fit (green) superposed onto the observed broad profile (red) of \mgiionly.}
		\label{fig:cornermgii}
	\end{figure*}

\begin{figure*}
		\centering 
		\includegraphics[width=0.48\textwidth]{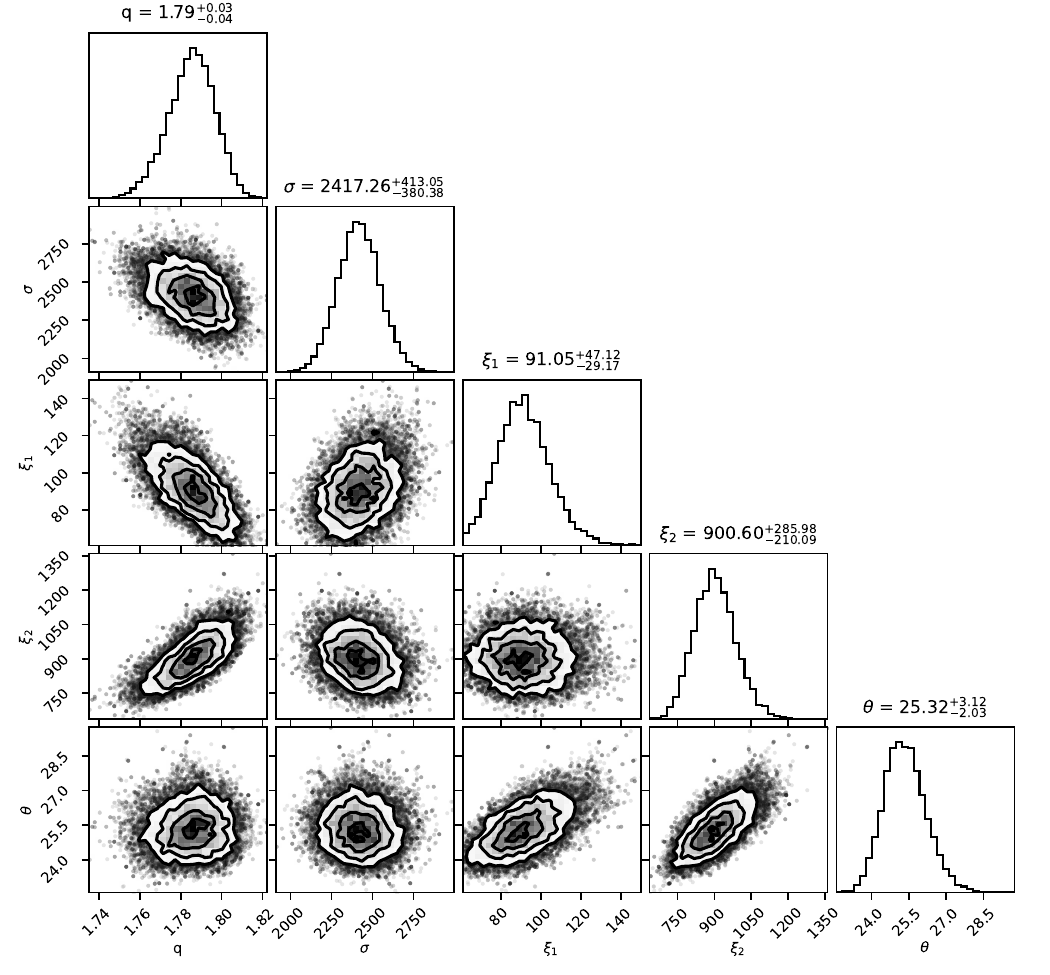}
		\includegraphics[width=0.48\textwidth]{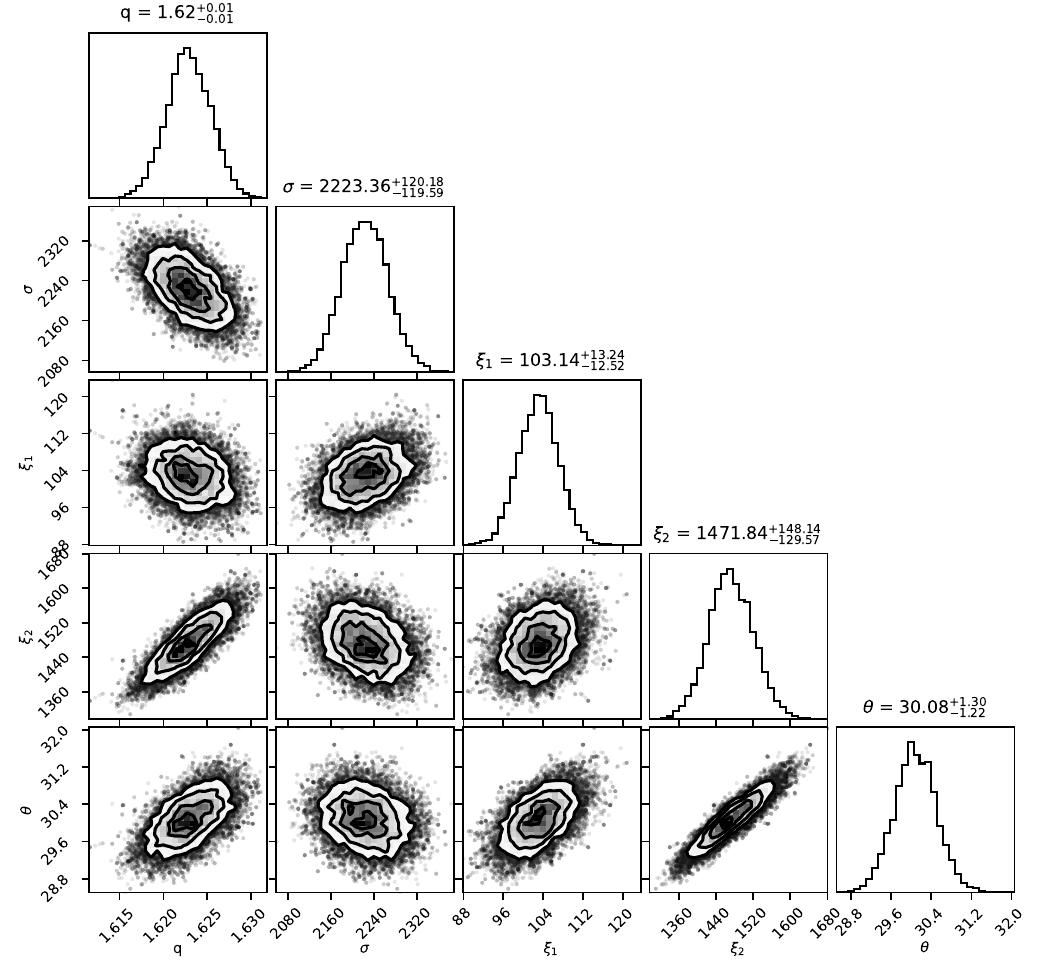}
		
		\caption {Corner plots for \hb\, (left panel) and \ha\, (right).}
		\label{fig:cornerhbha}
	\end{figure*}

\noindent In our case, the two sets of priors provide the same results with negligible differences between them and being well within the posterior distributions. Fig. \ref{fig:cornermgii} shows the corner plot (left panel) and 250 randomly selected posterior solutions (right) for the AD model (green lines) over the broad double-peaked profile (red) of \mgiionly. The corner plot shows the covariances and posteriors of the five parameters for the fit, in which at the top of each posterior distribution is displayed the median value of each parameter and errors measured as percentiles at the 0.135\% and 99.865\% levels (which correspond to $3\sigma$ errors in a normal distribution). In Fig. \ref{fig:cornerhbha} are represented the corner plots for \hb{} (left panel) and \ha{} (right). Table  \ref{tab:tabbayesian} contains the final measurements of the five model parameters for \mgiionly\, (Col. 2), \hb\, (Col. 3) and \ha\, (Col. 4) and the corresponding uncertainties.
	
	\begin{table}
		\centering
		\caption{Summary of the output parameters from the Bayesian inference model fittings.}
		\label{tab:tabbayesian} 
		\scalebox{1.}{
			\begin{tabular}{@{\extracolsep{2.pt}}l r c r c r}   
				\hline\hline \rule{0pt}{2ex}  
				& \multicolumn{1}{c}{\mgiionly} & & \multicolumn{1}{c}{H$\beta$} & & \multicolumn{1}{c}{H$\alpha$}\\[0.5ex]
				%\cline{2-3} \cline{5-6} \cline{8-9}
				%\vspace{0.1cm}
				\hline\rule{0pt}{2ex} 
				Parameters & \multicolumn{1}{c}{Median} & & \multicolumn{1}{c}{Median} & &\multicolumn{1}{c}{Median} \\
				
				\multicolumn{1}{c}{(1)} & \multicolumn{1}{c}{(2)} & & \multicolumn{1}{c}{(3)} & & \multicolumn{1}{c}{(4)} \\[0.5ex] 
				\hline \rule{0pt}{3ex} 
				q & 1.64$^{+0.01}_{-0.01}$ &  & 1.79$^{+0.03}_{-0.03}$ &  & 1.62$^{+0.01}_{-0.01}$ \\[1.5ex] 
				$\sigma^{a}$ & 1747$^{+163}_{-139}$ & &   2445$^{+404}_{-368}$ & & 2223$^{+120}_{-120}$ \\[1.5ex]
				$\xi_{1}(r_{g})$ & 131$^{+24}_{-22}$ & & 105$^{+43}_{-36}$ & & 103$^{+13}_{-13}$ \\[1.5ex]
				$\xi_{2}(r_{g})$ & 1632$^{+236}_{-204}$ & & 1013$^{+340}_{-242}$ & & 1472$^{+148}_{-130}$ \\[1.5ex]
				$\theta$(degrees) & 32.6$^{+2.4}_{-2.0}$ & & 26.7$^{+3.1}_{-2.5}$ & & 30.1$^{+1.3}_{-1.2}$  \\[1ex]
				\hline                     
			\end{tabular}
		}
		{\justify\textbf{Note}: {$^{a} \sigma$ is in units of \kms. The uncertainties in each parameter correspond to the confidence interval corresponding to 0.13\% and 99.87\% of the distribution of obtained values shown in the corner plots (Figs. \ref{fig:cornermgii} and \ref{fig:cornerhbha}).\par}}
	\end{table}
	
	\subsection{Empirical fittings}
	\label{sec:empirical}
	
	\subsubsection{{\mgiionly. \hb{} and \ha}}
	\label{sec:empiricalhb}
	
A non-linear multicomponent fitting procedure by using 
 {\tt specfit} was implemented to get a global fitting in each spectral range. We followed the same method described in \citetalias{2023MNRAS.525.4474M}. The {\tt{specfit}} fittings were performed twice. First, to represent the observed broad profile of each double-peaked line of interest (i.e., for \hb, \mgiionly\ and \ha) we assumed two Gaussian broad components (BC) of the same FWHM (corresponding to the blue and red components of the double-peaked profile) and a very broad component (VBC), which allowed us to isolate the broad profile in each spectral region after removing other components and to obtain the AD model parameters from the Bayesian fit, as explained in Sec. \ref{sec:bayesian}. In the second fitting, the resulted AD model was introduced to represent the broad profile instead of the initial 2BC + VBC. 
 
 In these fittings, we also included a power law local continuum \citep{1978Natur.272..706S}, an \feii\ template for modeling the \feiiopt\ and \feiiuv\ using the templates from \citet{2009A&A...495...83M} and \citet{2008ApJ...675...83B} respectively, and an additional UV Balmer continuum, found to be important at $\lambda$\,<\,3646\AA\, in the region of \mgiionly\ \citep{2014AdSpR..54.1347K, 2017FrASS...4....7K}. In addition, to represent the non-disk emission components, we include Gaussian profiles to fit the narrow and semi-broad components (NC + SBC) from the NLR.  All the narrow lines were assumed to have roughly the same width and shift values.	
 
 In the \hb\ region, the overall fitting was done in the wavelength range from 4400\AA\,--\,5300\AA. This includes, apart from the AD broad component, the \oiiiopt\ and \heii\ lines in addition to the \hb\ NC + SBC.  The \ha{} region final fit was done from 6100\AA\,--\,6900\AA. To get the best empirical fitting of the \ha\ + [NII]\ blend we include, together with the AD model output representing the broad profile, Gaussians  NC + SBC for \ha, \oi{} and for the doublets of \nii\, and \sii. A power law that defines the continuum and the \feii\ template, were also added.  The near-UV \mgiionly\ region was fitted in the wide wavelength range from 2600\AA\,--\,3800\AA, that includes the doublet NCs of \mgiionly{}, and {\sc{Oiii}}$\lambda$3133\AA, {\sc{[Oii]}}$\lambda$3727\AA\  and the two HILs of [NeV] at $\lambda\lambda$3346,3426\AA\AA. 
 
\begin{figure}
		\centering
		\includegraphics [scale = 0.40]{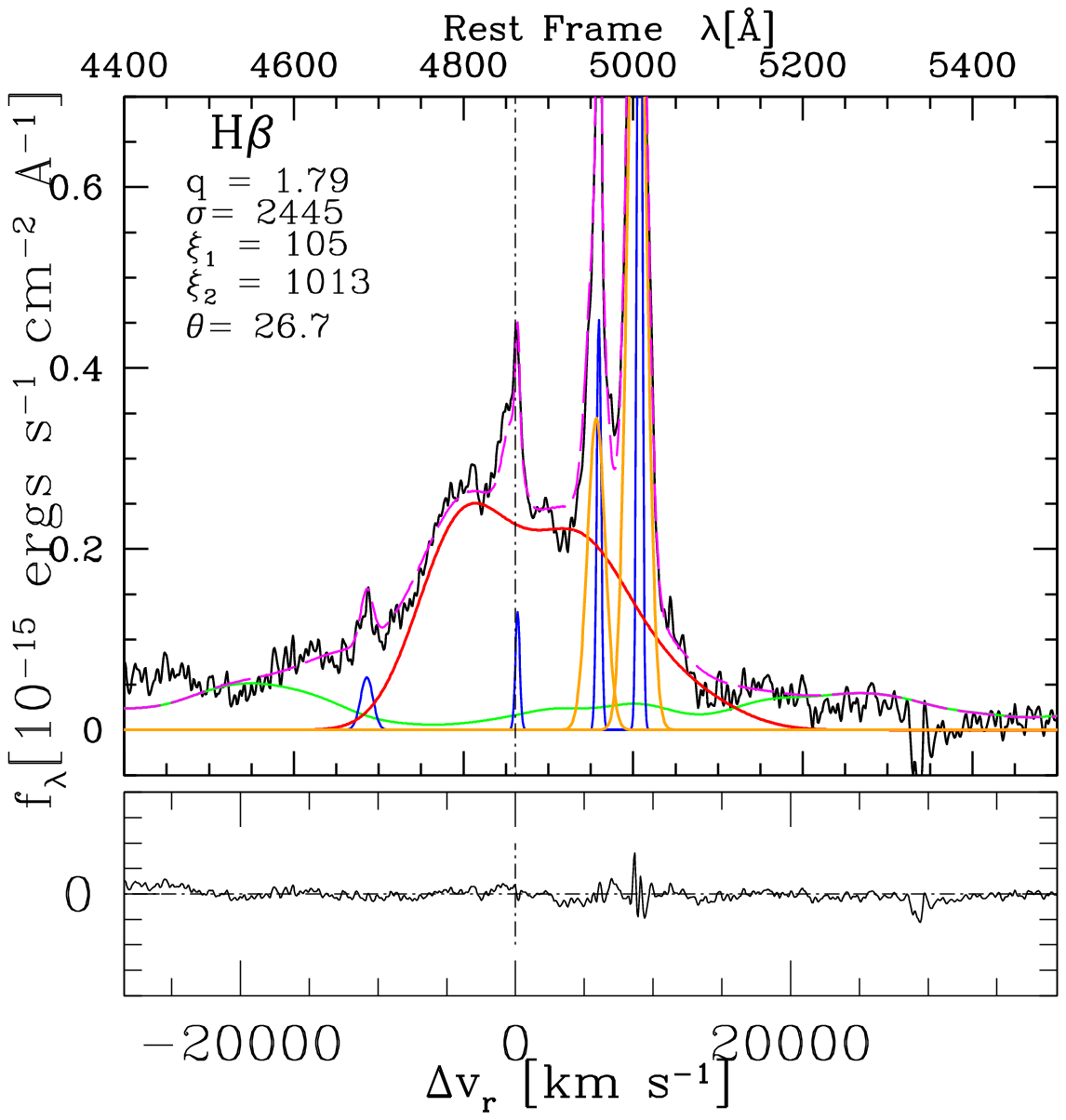}
		\caption{Multicomponent empirical {\tt specfit} analysis, including the AD model fitting result in the \hb\, region, after subtracting the continuum from the best fit. The upper abscissa is the rest-frame wavelength in \AA\ and the lower one is in radial velocity units. The vertical scale corresponds to the specific flux in units of 10$^{-15}$ergs\,s$^{-1}$\,cm$^{-2}$\,\AA$^{-1}$. The emission line components used in the fit are \feii\ (green), the broad AD model representing the fit for the broad double-peaked profile (red line), SBC (orange), and NC (blue). The black continuous line corresponds to the rest-frame spectrum. The dashed magenta line shows the final fitting from {\tt specfit}. The dot-dashed vertical lines trace the rest-frame wavelength of \hb. The lower panel shows the residual of the empirical fit.}
		\label {fig:hb}
	\end{figure}

\begin{figure}
		\centering
		\includegraphics [scale = 0.40]{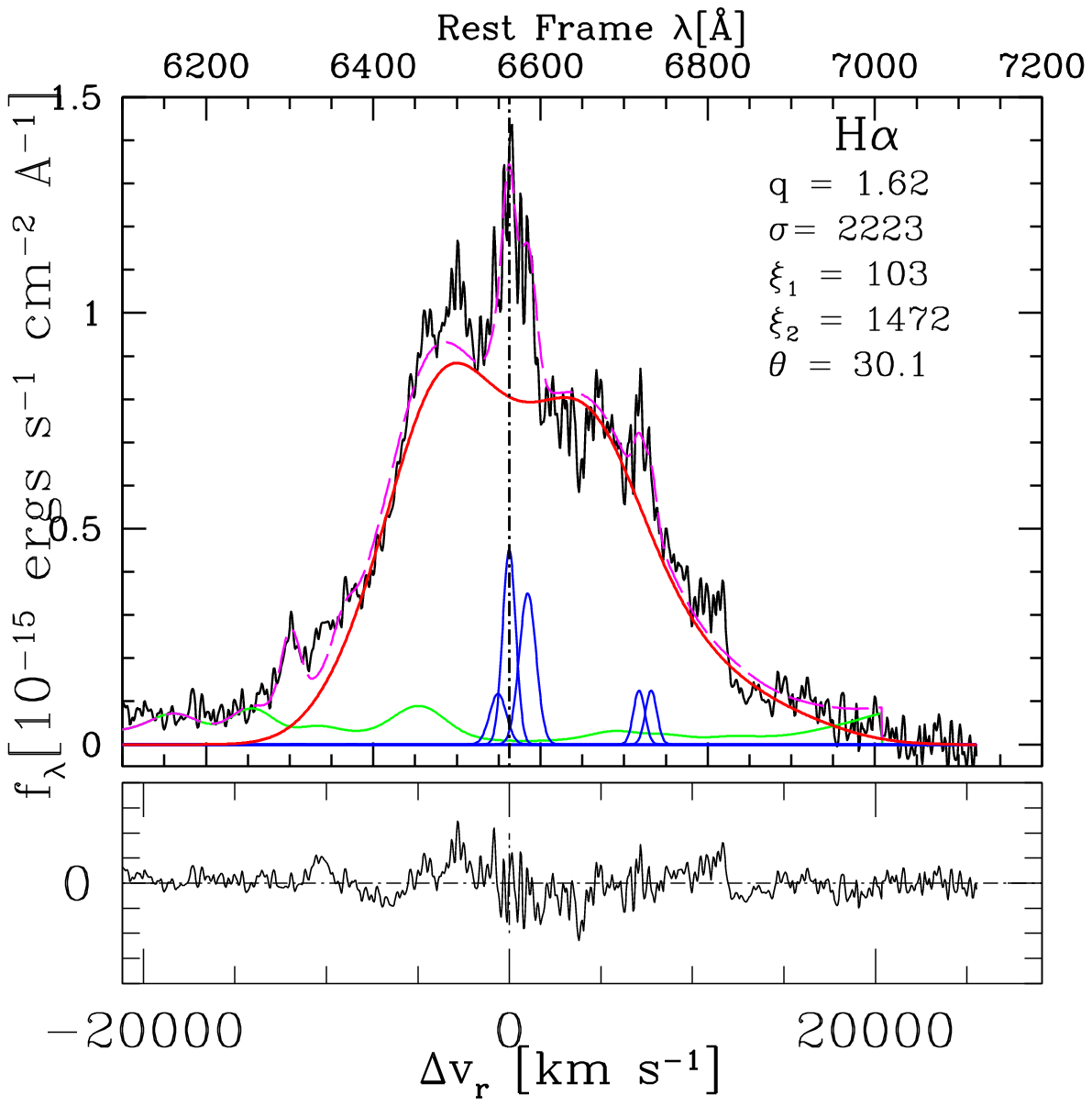}
		\caption{Multicomponent empirical analysis and AD model fitting results in the \ha\ region. The description is as in Fig. \ref{fig:hb}.
		}
		\label {fig:ha}
	\end{figure}

\begin{figure}
		\centering
		\includegraphics [scale = 0.40]{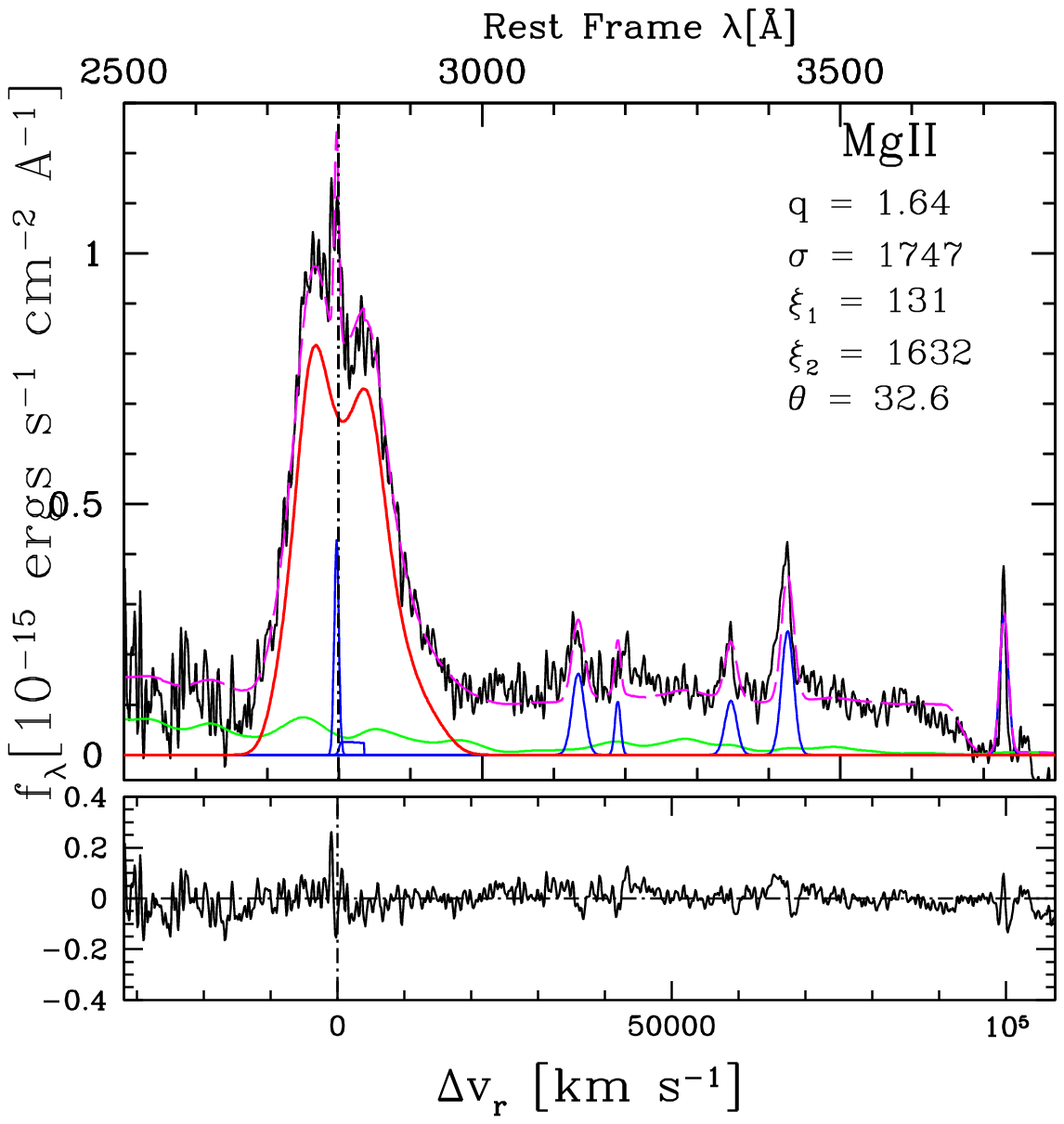}
		\caption{Multicomponent analysis and AD model fitting results in the \mgiionly\ region. The description is as in Fig. \ref{fig:hb}.}
		\label {fig:mg}
	\end{figure}

The resulting {\tt specfit} fits, including the AD model for the observed broad profiles as well as the non-disk emission components with a minimum $\chi^{2}$ in the regions of \hb, \ha{} and \mgiionly{} are shown in Figures \ref{fig:hb}, \ref{fig:ha} and \ref{fig:mg}, respectively, on a rest-frame wavelength scale. 
	
	% Fit Hbeta region

 %  Fit Halpha region
 
	% Fit MGII region
	
	\subsubsection {Civ$\lambda$1549 and Ciii]$\lambda$1909} 
	\label{subsec:empiricalc}
	
	Theoretical predictions and studies on double-peaked profile sources indicate that the disk emits predominantly LILs (e.g., Balmer lines, \mgiionly\, and \feii). High-ionization species, such as \civ\ may not be produced and may lack double-peaked profiles even when the LILs do show them \citep[e.g.,][]{1990A&A...229..292C, 1994ApJS...90....1E, 1998AdSpR..21...33E, 2009ApJ...704.1189T}. 
	
\begin{figure}
		\centering
		\includegraphics [scale = 0.4]{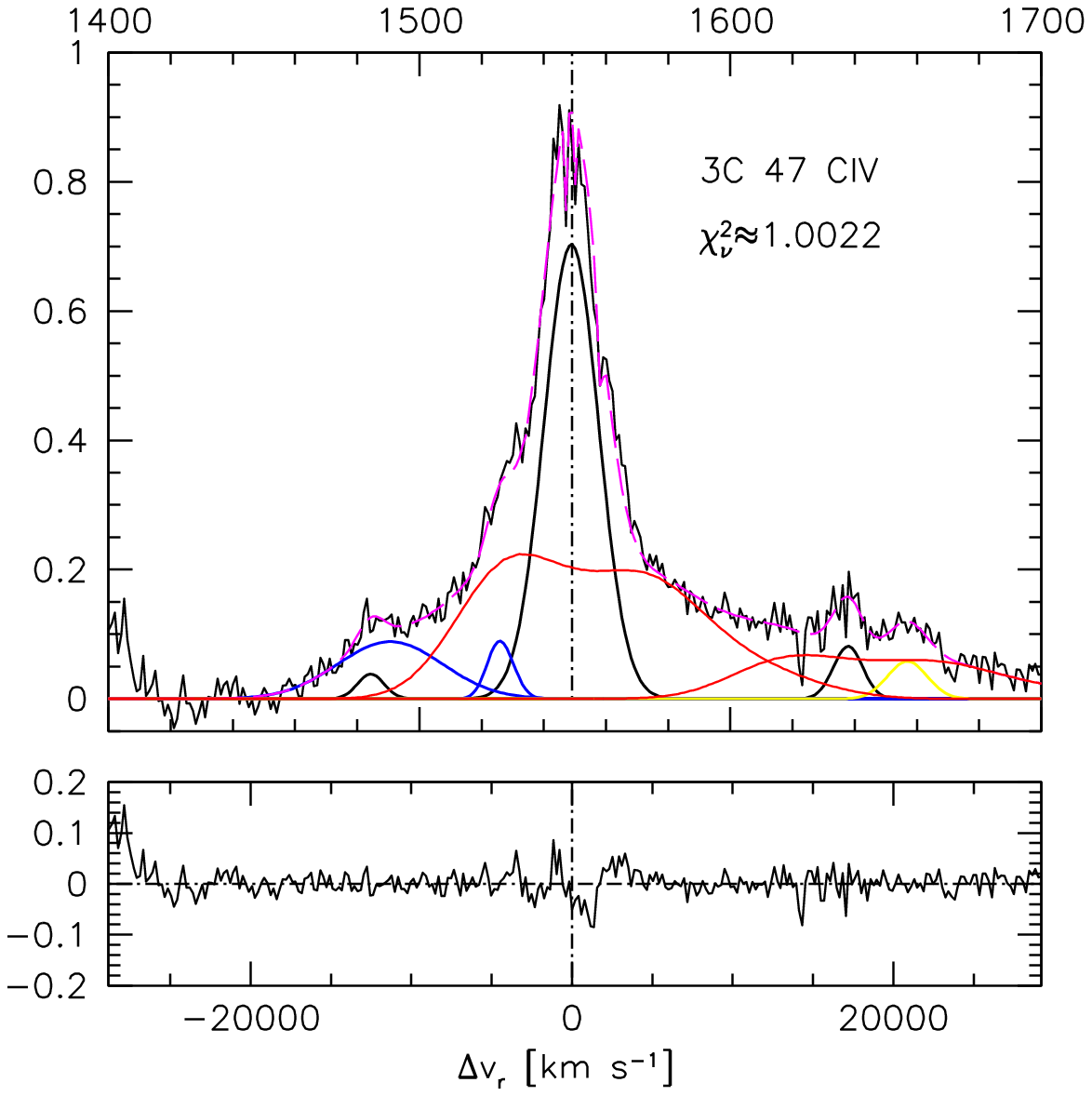}
		\caption{ Multicomponent empirical analysis and AD model fitting results for \civ\ and HeII$\lambda$1640. The black line corresponds to the rest-frame spectrum,  while the emission line components used in the empirical fit are blueshifted components (blue) and BC (black). The red line represents the model fitting result for \civonly\ and HeII$\lambda$1640 by using the AD model fitting parameters of \hb. The dashed magenta line shows the final model fitting from {\tt specfit}. The lower panel shows the residuals.}
		\label{fig:c4}
	\end{figure}

\begin{figure}
		\centering
		\includegraphics [scale=0.4]{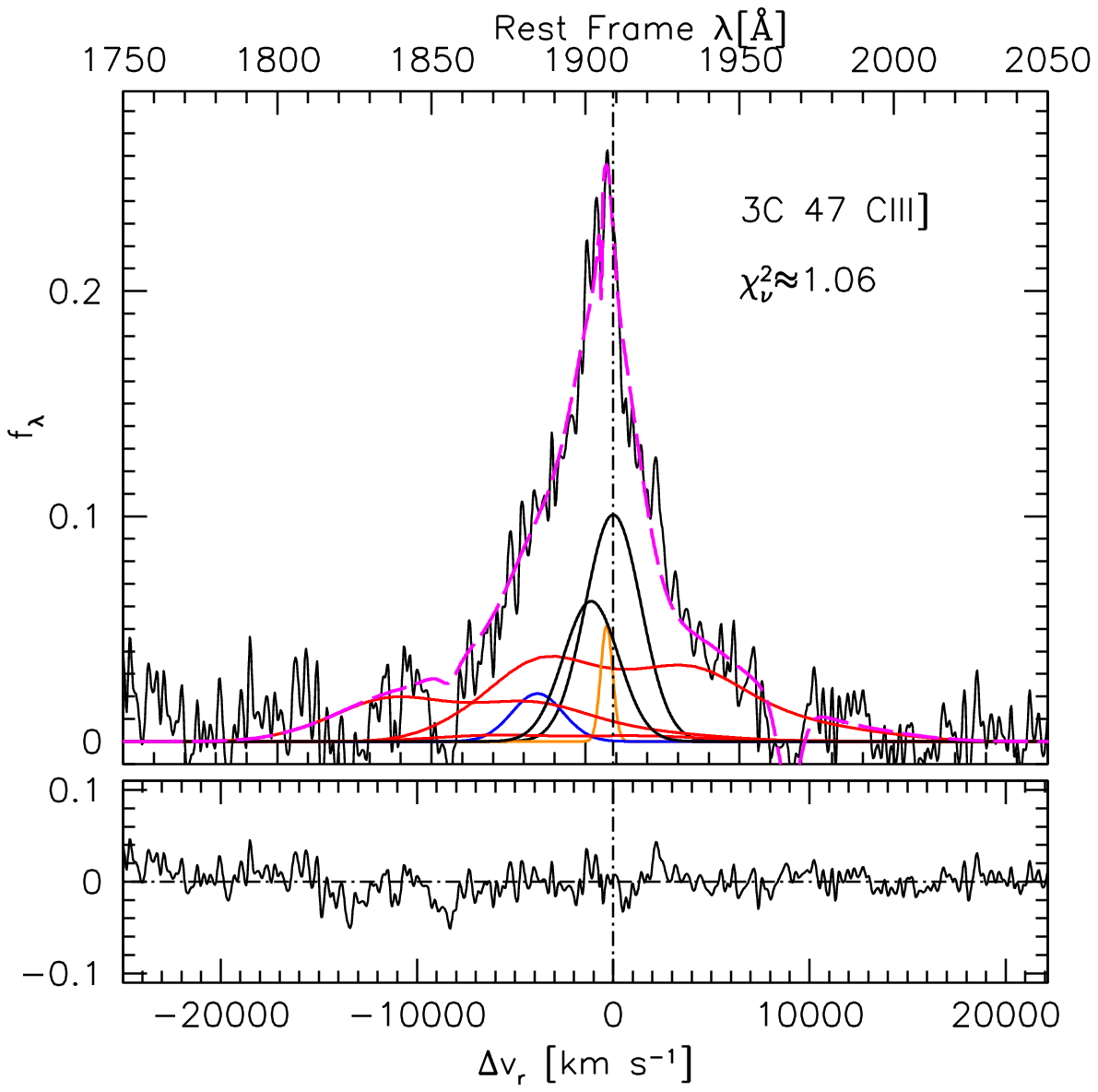}
		\caption{ Multicomponent non-linear empirical {\tt specfit} and AD model fits for \ciii\ blend following the same approach employed for \civ.  The axes and the components are as in Fig. \ref{fig:c4} except for the red line that represents the AD model fitting result for \ciii, \siiii, and the sum of the two \aliii\ components by using the parameters appropriate for the \mgiionly. The resulting \siiii\ is barely visible. The black lines trace the BC components of \ciii\ and \siiii.  
			The dot-dashed vertical line identifies the rest-frame wavelength of \ciii. 
		}
		\label{fig:c3}
	\end{figure}

	In this work, in addition to the LILs, we also fitted the available HIL: \civ, HeII$\lambda$1640, and \ciii. 
  The \civ\ line blend was fit in the spectral window 1420\AA\ -- 1705\AA\ and modeled by the contribution of the AD, along with prominent additional components including a symmetric Gaussian +  two blueshifted \civ\ components associated with a failed wind scenario, {\sc{Niv}}$\lambda$1483, 5 absorption lines eating away the NLR contribution to \civ, a  blueshifted SBC for HeII$\lambda$1640, and {\sc{Oiii}}$\lambda$1663. For the \ciii\ line profile, we considered the spectral window 1850\AA\ –1920\AA\ that includes two {Al\sc{iii}}$\lambda\lambda$1857,1862,  {Si\sc{iii]}}$\lambda$1892, \ciii\ and 2 absorption lines. We also assume a NC for \ciii. The resulting fits for the  \civ\ and \ciii\ regions are shown in Figures \ref{fig:c4} and \ref{fig:c3},  respectively.  A  discussion about the assumptions and the resulting fitting is presented in Sect. \ref{sec:failed}. 
	
\section{Results}
\label{sec:results}
\subsection{AD model parameters}
	\label{sec:mod}
	
The resulting quantitative measurements of the AD model,  i.e. $\{q, \sigma, \xi_{1}, \xi_{2}, \theta\}$, determined by following the approach described in Sect. \ref{subsec:model}, for \mgiionly, \hb, and \ha\ regions are reported in Table \ref{tab:tabbayesian}. The line emitting portion of the disk 
was found to be an annulus within $\xi_{1}\sim 100 r_\mathrm{g}$ and $\xi_{2}\sim 1000  r_\mathrm{g}$, with the largest values  $\xi_{1} \approx  131 r_\mathrm{g}$ and  $\xi_{2} \approx 1632 r_\mathrm{g}$\ estimated for \mgiionly. The ratio of outer to inner radius $\xi_{2}$/$\xi_{1}$ is $\approx$ 10, 12 and 14  for \hb, \mgiionly\ and \ha, respectively.  A larger value of $\xi_{2}$/$\xi_{1}$ ratio will bring the blue and red peaks close together and will form a single peak. 
	
	 We found a range of inclination angles relative to the line of sight between the two radii.  The lowest inclination of $\theta \approx$ 26.7$^{\circ}$ was found for \hb\ line and the largest inclination of $\theta \approx$ 32.6$^{\circ}$ for \mgiionly. The average value of ${\theta}$ is $\approx$ 29.8 $\pm$ 3.0 or $\sim$ 30\,$^{\circ}$. 
	
	The emissivity index ranges between 1.62 (\ha) -- 1.79 (\hb). The factor that represents the turbulent motion and is assumed to be the cause of local line broadening was represented by $\sigma/\nu_{0}$. We found the largest value of $\sigma/\nu_{0} = 8.156 \times10^{-3}$ for the \hb\ and the lowest for \mgiionly, of $\sigma/\nu_{0} = 5.827\times10^{-3}$. This value of $\sigma/\nu_{0}$ corresponds to a velocity dispersion $\sigma$ that is higher in \hb, with a value of 2445 \kms\ and lower in \mgiionly, with a value of 1747\kms. These values are comparable to the values obtained from  AD model fitting of other double peaked AGN \citep[e.g., ][]{1998AdSpR..21...33E,2003AJ....126.1720S}.  
	
	\begin{table*}
		\hspace{-1cm}
		\caption{Measurements of the full broad profile and the narrow (NC) + semi-broad component (SBC) of the three analyzed lines.}
		\label{tab:tab3}
		\setlength{\tabcolsep}{1pt}
		\scalebox{0.95}{
			\begin{tabular}{@{\extracolsep{3.5pt}}l  cccc c c c c cccccc c ccc   @{}} 
				\hline\hline 
				\rule{0pt}{2ex}
				% inserting double-line
				&\multicolumn{8}{c}{Full broad profile (FP)} && & &\multicolumn{3}{c}{NC + SBC component}  \\[0.5ex]
				\cline{2-9}\cline{13-15}
				% \noalign{\smallskip}
				Line & \multicolumn{1}{c}{Flux} & \multicolumn{1}{c}{FWHM} & \multicolumn{1}{c}{$c(\frac{1}{4}$)} & \multicolumn{1}{c}{$c(\frac{1}{2}$)} & \multicolumn{1}{c}{$c(\frac{3}{4}$)} & \multicolumn{1}{c}{$c(\frac{9}{10}$)} & \multicolumn{1}{c}{A.I.} & \multicolumn{1}{c}{K.I.}&f$_{\lambda}^{a}$&\feii$^{b}$& & Flux & Peak & FWHM \\
				%\hline
				(1) & (2)& (3) & (4) &(5) &(6)&(7)&(8)&(9)&(10)&(11)& &(12)&(13)&(14) \\
				\hline
				\noalign{\smallskip}
				\mgiionly\ & 116.7 & 15000$\pm$700 & 910$\pm$440 & 400$\pm$250 & 150$\pm$260 & -3110$\pm$390 & 0.43$\pm$0.34 & 0.60$\pm$0.04 & 1.94 & 41.9 & & 3.42$^{c}$ & 60 & 800\\[0.3ex]
				H$\beta$& 68.3 & 16600$\pm$800 & 1650$\pm$460 & 830$\pm$290 &  300$\pm$340 & -2360$\pm$720 & 0.42$\pm$0.27 & 0.59$\pm$0.04 & 0.73 & 9.8 & & 3.37 & 30 & 609 \\[0.3ex]
				H$\alpha$ & 313.0 & 15500$\pm$800 & 1020$\pm$460 & 430$\pm$270 & 110$\pm$300 & -430$\pm$2480: & 0.38$\pm$0.28 & 0.58$\pm$0.04 & 0.54 & 15.9 & & 8.86 & -90 & 840\\ [0.2ex]
				\hline 
			\end{tabular}
		}
		{\small{\justify\textbf{Notes}: {The full broad profile is considered to be the model fitting result that represents the broad profile of the observed spectra without including the NCs. Cols. 2, 11, and 12 are in units of 10$^{-15}$ergs\,s$^{-1}$\,cm$^{-2}$; FWHM, centroids and peak velocities in \kms;    $^\mathrm{a}$ Col. (10) corresponds to the power law continuum flux at  $\lambda$3000\AA, $\lambda$5100\AA, and $\lambda$6200\AA\  for \mgiionly, \hb\ and \ha\ regions respectively, in units of 10$^{-15}$ergs\,s$^{-1}$\,cm$^{-2}$\,\AA$^{-1}$. $^\mathrm{b}$ Fe{\sc ii} integrated flux in the range from  2200\AA\,--\,3090\AA,  4434\AA\,--\,4684\AA, and 6100\AA\,--\,6400\AA\ for \mgiionly, \hb\ and \ha\, respectively (in units of 10$^{-15}$  
        ergs\,s$^{-1}$\,cm$^{-2}$).  $^{(c)}$ Intensity of NC is for  the \mgiionly\, blue line of the doublet (a ratio of 1.5 was assumed for the NCs of the  
		the doublet $\lambda$2796.35\AA\, and $\lambda$2803.53\AA.) \par}}}
	\end{table*}
	
	\begin{table*}
		\centering
		\hspace{-1cm}
		\caption{Measurements of the full broad profile and the narrow + semi-broad component of the FUV lines.}
		\label{tab:tabuv}
		\setlength{\tabcolsep}{2pt}
		\scalebox{0.95}{
			\begin{tabular}{@{\extracolsep{3pt}}l  cccc c c c c cccccc c ccc   @{}} 
				\hline\hline % inserting double-line
				&\multicolumn{8}{c}{Full broad profile (FP)} && \multicolumn{3}{c}{NC + SBC component}  \\
				\cline{2-9}\cline{11-13}
				Line & \multicolumn{1}{c}{Flux} & \multicolumn{1}{c}{FWHM} & \multicolumn{1}{c}{$c(\frac{1}{4}$)} & \multicolumn{1}{c}{$c(\frac{1}{2}$)} & \multicolumn{1}{c}{$c(\frac{3}{4}$)} & \multicolumn{1}{c}{$c(\frac{9}{10}$)} & \multicolumn{1}{c}{A.I.} & \multicolumn{1}{c}{K.I.}&f$_{\lambda}^{a}$&  Flux & Peak & FWHM \\
				%\hline
				(1) & (2)& (3) & (4) &(5) &(6)&(7)&(8)&(9)&(10)&(11)&(12)&(13)\\
				\hline
				\hline
				{\sc Civ}\ & 387.0  & 4972 & -345 & -125 & -100 & -93 & -0.06 &  0.33 & 2.37 &  9.45 & -4500$^\mathrm{b}$ & 1850 \\
				He{\sc ii}\ & 71.9 & 3758 & 731 & -678 & -525 & -514 & 0.15 &  0.12 & 2.37   & \ldots & \ldots & \ldots  \\
				Al{\sc iii}  & 92.7 & 14681 & 1000 & 355 & 113 & -452 & 0.15 &  0.59 & 2.97  & \ldots & \ldots & \ldots  \\
				C{\sc iii}] & 373.9 & 5553 & -817 & -687 & -267 & -188 & -0.15 &  0.36 & 2.97  & 4.8 & -200/--500$^\mathrm{c}$ & 910 \\
				\hline 
			\end{tabular}
		}
		{\small{\justify\textbf{Notes}: { Cols. 2 and 11 are in units of 10$^{-15}$ergs\ s$^{-1}$cm$^{-2}$; $^\mathrm{a}$ Col. (10) yields the continuum specific flux at 1350\AA\ and 1750\AA, for the \civ\ + HeII$\lambda$1640, and the 1900\AA\, blend regions respectively, in units of 10$^{-15}$ergs\ s$^{-1}$cm$^{-2}$\AA$^{-1}$. $^\mathrm{b}$\,Semi-broad blueshifted emission.  $^\mathrm{c}$ Reference wavelength 1906.7\AA\, and 1908.7\AA, respectively. \par}}}
	\end{table*}
	
	\subsection{Empirical fitting results}
 
 We employed a comprehensive parameterization of the broad profiles involving total flux, FWHM, and centroids at different fractional line intensities ($c(i/4)$ for i = 1, 2, and 3) and 9/10, $c (9/10)$ (a proxy to the line peak), as well as asymmetry (AI) and kurtosis (KI) indices \citet{2010MNRAS.403.1759Z}. 
The 3C 47 parameters are reported in Table \ref{tab:tab3} for \mgiionly, \hb\, and \ha.   
The broad line fluxes (Col. 2) and the specific fluxes of the power-law continuum (Col. 10), along with the FWHM (Col. 3) were used for the computation of the accretion parameters ( see Sect. \ref{sub:accr}).

\object{3C 47} shows very large line width with FWHM of $\gtrsim$ 15000 \kms. The \hb\ disk line profile is among the broadest, with FWHM of 16600 \kms. The centroid velocity measurements at different fractional intensities show a predominance of shifts to the red, being more pronounced towards the line base, with c(1/4)\,$\approx$\,1650$\pm$460\,\kms.
Within the uncertainty level, the redward asymmetry appears consistent across all three lines, with \ha{} showing a slightly more symmetric distribution compared to \hb{} and \mgiionly.

The flux of the \feii\ emission integrated for \mgiionly, \hb, and \ha\ regions are reported in Col. 11 of Table \ref{tab:tab3}.  With an intensity ratio of approximately 0.14 for {\rfeopt= I(\feiiq)/I(\hb)}, \object{3C 47} falls within the realm of extreme Population B on the optical plane of the quasar main sequence (MS), defined by the relationship between \hb\ FWHM and \rfeopt\ \citep{2002ApJ...566L..71S}. More precisely, its \hb\ FWHM of $\approx$16,000 \kms\ places it in the B1$^{++}$ ST according to the classification scheme of \citet{2002ApJ...566L..71S}. It is noteworthy that this ST comprises a minimal fraction (less than 1\%) of optically selected quasars with weak or undetectable \feii\ emission \citep{2013ApJ...764..150M}. \citet{2019A&A...630A.110G} revealed a higher prevalence of irregular/multiple peaked profiles in this ST and suggested a higher prevalence for binary SMBH candidates (see Sect. \ref{sec:alter}).

 Cols. 1 -- 9 of Table \ref{tab:tabuv} report the full broad profile parameters for the far-UV (FUV) lines, mirroring the organization of Table \ref{tab:tab3} along with the specific flux (Col. 10). A key result is that the FWHM \civ\ $\ll$ FWHM \hb, which will be discussed 
 in more detail in Sect. \ref{sec:failed}.
Highly blueshifted semi-broad features (blue lines in Figs. \ref{fig:c4} and \ref{fig:c3}) visible for both \civ\ and \ciii\  are probably part of a continuum of outflowing gas from the outer BLR to the inner NLR. Their fluxes -- much lower than the other components -- may indicate that outflows may have little effect over the integrated line profiles (see Sect. \ref{sec:failed} for more detail).

In addition to the parameters of the broad profiles, we report the flux, the peak shift, and the FWHM for NCs and SBCs of identified emission lines from the {\tt specfit} analysis (Tables \ref{tab:tab3} and \ref{tab:tabuv}). Individual NCs are near the rest-frame, with a broad FWHM of 840 \kms\ for \ha\ and a slightly lower value for \hb{} and \mgii. The \civ\ NC is absorbed by narrow lines close to the rest-frame. 

 Finally, Table \ref{tab:tab31} reports total intensity, peak shift, and FWHM for additional narrow and semi-broad emission lines identified in the four spectral ranges (\mgii, \hb, \ha, and \civ). 

		\begin{table}
			\centering
			\caption{Narrow lines and narrow line components.}
			\begin{tabular}{lrccc } 
				\hline\hline
				\noalign{\smallskip}
				Spectral Range & \multicolumn{1}{c}{Line} & Flux & Shift&FWHM \\[0.3ex]
				\hline
				\multirow{3}{4em}{\mgiionly} & O III$\lambda$3133 &3.49&40& 1940 \\ 
				& He I$\lambda$3188 &1.05 &60& 880\\ 
				&[Ne v]$\lambda$3426 &5.22 &-30&1740\\ 
				&[O II]$\lambda$3728 & 4.19&10&1140\\ 
				\hline
				\multirow{3}{4em}{\hb} & \oiiiseven & 8.86&-40&370 \\ 
				&\oiiiseven$^{a}$ & 27.05 &-220& 1490\\ 
				& \heii &1.06 &-70 &1100\\ 
				\hline
				\multirow{4}{4em}{\ha} & [O I]$\lambda$6302 &0.01&380& 570 \\ 
				&  [N II]$\lambda$6585&8.79 &-30&1070 \\ 
				& [Si II]$\lambda$6732& 2,14&-10 &720\\ 
				&[O I]$\lambda$6302$^{a}$ & 5.35&-60 &1310\\
				\hline
				\multirow{2}{4em}{\civ} &  N IV] $\lambda$1486& 4.1 & -420 & 2020 \\ 
				& O III] $\lambda$1663&  9.2 & --1100 & 2700 \\ 
				\hline
				\\[-0.8ex]
			\end{tabular}
			{\raggedright\textbf{Notes}: {$^{a}$ represents the flux, peak shift and FWHM of the SBC. Intensities are in units of 10$^{-15}$ ergs\,s$^{-1}$\,cm$^{-2}$.\par}}
			\label{tab:tab31}
			%\end{center}
		\end{table}
		
		\begin{table}
			\centering
			\caption{{\ Accretion parameters. }}
			\label{tab:tab4}
			\setlength{\tabcolsep}{1pt}
			\scalebox{0.9}{
				\begin{tabular}{@{\extracolsep{2pt}}l  cccccc@{}} 
					\hline\hline % inserting double-line
					\noalign{\smallskip}
					Line/continuum &  \multicolumn{1}{c}{$\log L $}& \multicolumn{1}{c}{ $\log  r$} &  Method&  \multicolumn{1}{c}{$ M_\mathrm{BH}$} &\multicolumn{1}{c}{$L_\mathrm{bol}/L_\mathrm{Edd}$} & Refs.
					\\
					%\hline
					(1) & (2)& (3) & (4)& (5) & (6) &\multicolumn{1}{c}{(7)}\\[0.2ex]
					\hline
					\noalign{\smallskip}
					L(\hb)$^\mathrm{a}$	&	43.33	&	17.70	&	Vir	  &	7.89E+09	&	0.036&	1  \\
					L(5100)	&	45.22	&	17.60	&	Vir	  &	6.26E+09	&	0.045&2 	         \\
					L(5100)	&	45.22	&	17.59	&	Vir	 &	6.08E+09	&	0.047	&3	 	       \\
					L(3000)	&	45.42	&	17.63	&	Vir	  &	5.42E+09	&	0.052&	4  	\\
					L(3000)	&	45.42	&	17.72	&	Vir	  &	6.65E+09	&	0.043&5	 	\\
					L(3000)	&	45.42	&	18.06	&	Vir	  &	1.46E+10	&	0.019&2	 	\\
					Average	&		&		&						&	6.46E+09	&	0.040	\\
					St.dev.	&		&		&						&	9.14E+08	&	0.012	\\
					&		&				&		&		\\
					L(\ha)	&	44.12	&	\ldots	&	SL	  &	8.04E+09	&	0.035 &	6 	   \\
					L(5100)	&	45.22	&	\ldots	&	SL	&	9.14E+09	&	0.031 &7	 	\\
					L(5100)	&	45.22	&	\ldots	&	SL	  &	7.96E+09	&	0.036 &2 	    \\
					L(\hb)	&	43.43	&	\ldots	&	SL	  &	5.18E+09	&	0.055 &8	 	\\
					L(3000)	&	45.42	&	\ldots	&	SL		&	9.33E+09&	0.030 &4	 	\\
					L(3000)	&	45.42	&	\ldots	&	SL	&	8.31E+09	&	0.034 &9 		\\
					L(\mgiionly) 	&	43.69	&	\ldots&	SL	 &	1.23E+10	&	0.023 &	7   \\
					L(3000)$^\mathrm{b}$	&	45.42	&	\ldots	&	SL	 &	5.34E+10	&	0.005 &2\\
					Average	&		&		&						&	8.61E+09	&	0.035	\\
					St.dev.	&		&		&						&	2.12E+09	&	0.010	\\
					\hline
				\end{tabular}
			}
			{\justify{\textbf{Note}: Col. 2 in ergs\ s$^{-1}$, Col. 3 in cm,  and Col. 4 in M$_\odot$. $^\mathrm{a}$For the same flux, the luminosity has been scaled to $H_0 = 75$ \kms\ Mpc$^{-1}$\ as used in the \citet{2004A&A...424..793W} relation. $^\mathrm{b}$\mbh\ and \lledd\ excluded from average and standard deviation as they are outliers above a $3\sigma$\ confidence level.  References: (1) \citealt{2004A&A...424..793W}; (2) \citealt{2023arXiv230501014S}; (3) \citealt{2013ApJ...767..149B}; (4) \citealt{2012MNRAS.427.3081T}; (5) \citealt{2019FrASS...6...75P}; (6) \citealt{2011ApJS..194...45S}; (7) \citealt{2006ApJ...641..689V}; (8) \citealt{2005ApJ...630..122G}; and (9) 	\citealt{2009ApJ...699..800V}.  \par}}
		\end{table}
		
		\section{Discussion}
		\label{sec:disc}
			
		\subsection{Accretion parameters}
		\label{sub:accr}
		
Black hole masses (\mbh) are computed from the virial equation applied in the form of a scaling law (SL), where \mbh\ has been assumed to be proportional to the square of the line width \citep[][]{2006ApJ...641..689V}, and the BLR radius has been estimated from its correlation with continuum or luminosity of different emission lines \citep{2012ApJ...753..125S,2013ApJ...767..149B}. Alternatively, the line width can be raised to a power different from the second \citep{2012ApJ...753..125S}. This approach has been extensively applied in previous studies, including large samples of quasars from the Sloan Digital Sky Survey (SDSS) \citep[e.g., ][and references therein]{2017ApJS..228....9K}.

 An important input parameter does not explicitly appear in the SLs: the viewing angle ($\theta$) between the line-of-sight and the axis of symmetry of the AGN (i.e., the AD axis or the radio axis). The double-peaked sources such as \object{3C 47} presumably have almost all of their low-ionization emission lines modeled by a geometrically thin, optically thick disk, and it is, therefore, easy to associate a well-defined $\theta$\ with the line's FWHM \citep[for an alternative approach, see e.g.,][]{2007ApJ...668..721B}. 
		
		Taking into account the inclination angle derived from the disk profile analysis, the virial relation can be written as: 
		\begin{equation}
			M_\mathrm{BH}=\frac{R_\mathrm{BLR}\delta v_\mathrm{K}^{2}}{G} = f(\theta)\frac{R_\mathrm{BLR}\mathrm{FWHM}^{2}}{G}
			\label{eq:vir}
		\end{equation}
		where $\delta v_\mathrm{K}$\ is the Keplerian velocity of the line emitting gas at the radius of the BLR ($R_\mathrm{BLR}$). The virial factor $f(\theta)$\ connects $\delta v_\mathrm{K}$ to the observed FWHM and can be written as \citep[e.g.,][]{2008agn8.confE..14D,2018NatAs...2...63M,2018A&A...620A.118N}:
		\begin{equation}
			f(\theta)=\frac{1}{4[k^2 + \sin^{2}\theta]} ,  k=\frac{\sigma v_\mathrm{iso}}{\sigma v_\mathrm{K}}\approx 0.3
		\end{equation}
		$f \approx 0.735$ when $\theta \approx 30$. 

Table \ref{tab:tab4} reports the line or continuum identifications and the corresponding luminosity values (Cols. 1 and 2, respectively), the BLR radius estimated from its correlation with luminosity (Col. 3), the method used (either using Eq. (\ref{eq:vir}) above, or a SL linking directly mass to luminosity and line width, Col. 4), and the estimated \mbh\  (Col. 5) and the Eddington ratio (\lledd) (Col. 6) for several different SL (in Col. 4, with reference in Col. 7) using \hb, \mgiionly\ and \ha\ as virial broadening estimators.
  
The \mbh{} estimates based on Eq. (\ref{eq:vir}) are only marginally lower than the ones based on the SLs ($(6.5\pm 0.9) \times 10^9$ M$_\odot$, vs $(8.6\pm 2.1) \times 10^9$ M$_\odot$), and both  \mbh\ values confirm the nature of \object{3C 47} as an evolved system with a SMBH at the high end of \mbh\ distribution function \citep[e.g.,][]{2009ApJ...699..800V,2017FrASS...4....1F}.

  The \lledd\ has been computed from the \object{3C 47} spectral energy distribution (SED) as available in the NASA/IPAC Extragalactic Database (NED). The mid (MIR) and far infrared (FIR) emissions have been cut to avoid the inclusion of reprocessed radiation and unresolved emissions not associated with the AGN. The resulting bolometric correction factors are 25.59 from $\lambda f_\lambda$ 5100\AA\ and 16.39 from $\lambda f_\lambda$3000 \AA, considerably higher than in the case of optically selected RQ quasars, $\approx$ 10 -- 15 from $\lambda f_\lambda$ 5100\AA\ \citep{2006ApJS..166..470R, 2020ApJ...903...44P}. For 3C 47, the bolometric correction includes the accretion luminosity (i.e., from the AD and associated coron\ae), as well as the emission associated with the relativistic jet. A flat X-ray SED, expected to be due to synchrotron self-Compton (SSC, \citealt{1992ApJ...397L...5M}) emission, is producing a luminosity of $\log L_\mathrm{X} \approx$ 46.28 [erg s$^{-1}$],  which contributes to $\approx$ 40 \%\ of the bolometric luminosity $\log L \approx$ 46.64 [ergs s$^{-1}$]. For comparison, a Population A AGN SED (corresponding to the \citet{1987ApJ...323..456M} template, as implemented by the command {\tt table AGN} in CLOUDY)  with the same optical flux of \object{3C 47} would yield $\log L_\mathrm{X} \approx$ 45.24 vs. $\log L  \approx$ 46.32 [ergs s$^{-1}$] \citep[e.g.,][]{2012MNRAS.425..623L}. The \lledd\ estimated from the average of the three \mbh\ determinations from Eq. (\ref{eq:vir}) is \lledd\ $\approx$ 0.040 $\pm$ 0.012, inclusive of the accretion and non-thermal emission. The SSC emission may be beamed; however, beaming effects on the optical/UV synchrotron continuum are probably minor, as confirmed by the \mbh\ estimates based on continuum measurements that are consistent but slightly lower than the ones based on emission lines for which no beaming effect is expected.
  		
		\subsection{AD as the origin of double-peaked profile}
		\label{sec:accre}
		
   Two evidences support an illuminated AD as the main source of LILs (see Sects \ref{sec:bayesian} and \ref{sec:results}, and Table \ref{tab:tabbayesian}):
		\begin{enumerate}
			\item The two peaks are well separated, implying a disk of modest extension with $\xi_{2}$ $\sim 10^3$\,r$_\mathrm{g}$.
			\item The overall profile is asymmetric,  due to relativistic beaming (Doppler boosted peak) and gravitational redshift of the entire line (redshifted line base).
		\end{enumerate}
		The most remarkable finding is that the $\xi_{1}$\ should be as low as $\sim 100$\,r$_\mathrm{g}$,  implying a large shift because of gravitational and transverse redshift, $c \delta z_\mathrm{grav} \approx {3}/{2\,\xi_1} \approx 4500$ km s$^{-1}$. It is also remarkable that the predicted shape provides a good fit for the line red wings. \object{3C 47} is not an exception in this respect, as blazars also show red asymmetric disk profiles characterized by large shifts at the line base, consistent with emission from the innermost disk \citep[see e.g.,][]{2020ApJ...903...44P,2023Symm...15.1859M}. In the case of blazars, the orientation angle is however $\theta \lesssim 5$\ $^{\circ}$ and, consequently, the profiles appear single peaked\,   \citep{2011MNRAS.413...39D,2023Symm...15.1859M}.   
		
		The decreasing line width in the order \hb, \ha, and \mgiionly\ is associated with a difference in the emissivity-weighted radius of the three lines. The \mgiionly\ is emitted by a region of the disk weighted toward larger radii than \hb\ and \ha\ (see Table \ref{tab:tabbayesian}, Cols. 2, 3 and 4). This effect is seen in Population B sources where emission is dominated by a virial velocity field \citep[][\citetalias{2023MNRAS.525.4474M}]{2013A&A...555A..89M}, for which the line width is inversely proportional to the square root of the distance from the central continuum source, i.e., FWHM\,$\propto\ {1}/{\sqrt{r}}$. 
		
		In summary, the three LILs show profiles that are consistent with a relativistic flat AD, and their main features can be explained by the physical processes expected to occur for ionizing continuum radiation reprocessed by the disk. This is, however, not the case for HILs ( see Sect. \ref{sec:failed} for more detail).  
		
		\subsection{Consistency between radio and optical properties}
		\label{radio-opt}
		
		Assuming that the relativistic jet is co-axial with the disk,  i.e. the jet axis and the disk plane are perpendicular, an interesting issue is whether the orientation of the jet is consistent with the inclination of the AD. The radio morphology is double-lobed, with jets but no counterjet easily visible at 4.9 GHz \citep{1991ApJ...381...63F}. The jet/counterjet asymmetry implies the presence of a highly relativistic jet and, along with the double-lobe morphology, a significant misalignment between the jet axis and the plane of the sky as well as the line-of-sight. Superluminal motion has been detected by VLBI observations of \object{3C 47}, with an apparent superluminal speed of $\beta_\mathrm{app} \approx  5.3^{+1.3}_{-1.0} $ \citep{1993ApJ...417..541V}. Such high superluminal speed is incompatible with an angle as large as 30$^{\circ}$, as the upper limit to $\theta_\mathrm{jet}$ is $\approx$ 20$^{\circ}$ with  Lorentz factor ($\gamma) \approx 10$. 
		However, considering that the uncertainties in the superluminal speed translate into $\theta_\mathrm{jet} \approx 20^{+6}_{-3}$, the difference between  $\theta_\mathrm{jet}$ and $\theta_\mathrm{disk} \approx 30 \pm${3} is significant only slightly more than at $\approx 1 \sigma$\ confidence level. It is also important to remark that the \object{3C 47} $\beta_\mathrm{app}$\ might be overestimated, as this source is an outlier in the correlation between  $\beta_\mathrm{app}$\ and deboosted core-to-lobe ratio, in the sense that the \object{3C 47} $\beta_\mathrm{app}$ is much higher than expected for its core-to-lobe ratio \citep{1993ApJ...417..541V}.

		The jet-to-counterjet flux ratio offers an independent test of consistency for the angle $\theta_\mathrm{jet}$. The counterjet is not visible, but an upper limit to the flux is $\approx 4.2$\ mJy \citep{1994AJ....108..766B}, yielding a lower limit to jet-to-counterjet ratio $\approx 5.5$. This in turn implies that $\gamma \gtrsim 2.5$\ for $\theta \approx 29$. 
		
		Therefore, we conclude that the radio-derived angle between the line-of-sight and the jet, and between the line-of-sight and the disk axis, $\theta_\mathrm{jet}$ and $\theta_\mathrm{disk}$ are not significantly discordant, leaving open the possibility of slight bending of the pc-sized jet over scales $\sim 10^4 r_\mathrm{g}$\ \citep[][ and Section \ref{sec:alter}]{2018RAA....18..108Y}.

		\subsection{A failed wind signature}
		\label{sec:failed}
		
  The most striking result is the obvious difference between the profiles of the LILs and HILs. Unlike in the case of Population A sources where the \civ\ profile is usually broader and blueshifted in comparison to the Balmer lines,  for several Population B RL sources, we can explain the \civ\ profile as due to the disk profile in addition to the emission from a strong, narrower feature (FWHM $\approx 4000$ \kms) peaking at rest-frame and with a slightly asymmetric profile. What is the origin of this feature? 
		
As mentioned above, the interpretation of the "double peakers" \citep[\citetalias{1989ApJ...344..115C};][]{1994ApJS...90....1E,2003AJ....126.1720S} is that the LILs are produced exclusively due to the reprocessing by the gas of the AD. The contribution from the disc is expected in the general AGN population, but is believed to be masked by emission due to the gas surrounding the disk itself  \citep{2004A&A...423..909P,2009MNRAS.400..924B}. Therefore, double peakers are, in this interpretation, low radiators and sources with little gas left and with a disk truncated at $\xi_{2} \lesssim 10^3 r_\mathrm{g}$ \citep{2021Univ....7..484M}. In addition, jetted sources are X-ray bright and may induce a high-ionization degree (an over-ionization of the gas; \citealt{1995ApJ...451..498M,1997ApJ...474...91M}). In a line driven wind scenario, the ionic species yielding the main resonant transitions that absorb the continuum momentum are replaced by higher ionization species, and the gas cannot be efficiently accelerated anymore {\citep[e.g.,][and references therein]{1995ApJ...451..498M,2004ApJ...616..688P,2024MNRAS.527.9236H}}. 
This scenario may apply to \object{3C 47}. We consider the disk radii that correspond to linear distances of $\sim 10^{17} (r_\mathrm{g}$/100)\, cm,  assuming \mbh\ $\approx 6.5 \times 10^9\,  \mathrm{M}_\odot$, and an array of photoionization simulations assuming as input: (1) the SED of \object{3C 47} as derived from the NED data; (2) radii in the range $\log \xi_{2} \sim 17 - 19$ [cm]; (3) hydrogen density between $\log n \sim 10$ [cm$^{-3}$] and $\log n \sim 12$ [cm$^{-3}$], with steps of $0.5$ dex; (4) metallicity $Z$ at 0.1 and 1 $Z_\odot$; (5) 0 turbulence broadening.   
		
\begin{figure*}
			\centering
			\includegraphics [scale = 0.3 ]{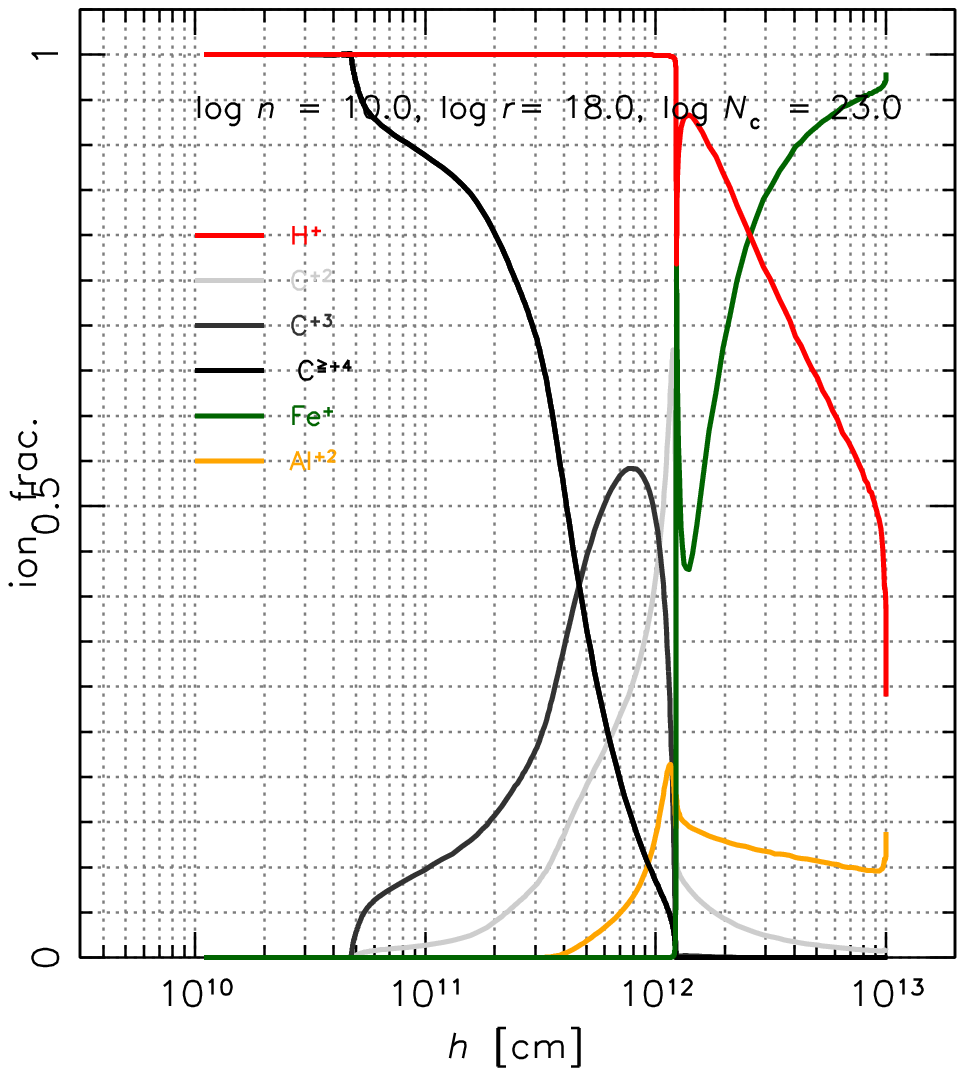}
			\includegraphics [scale = 0.3 ]{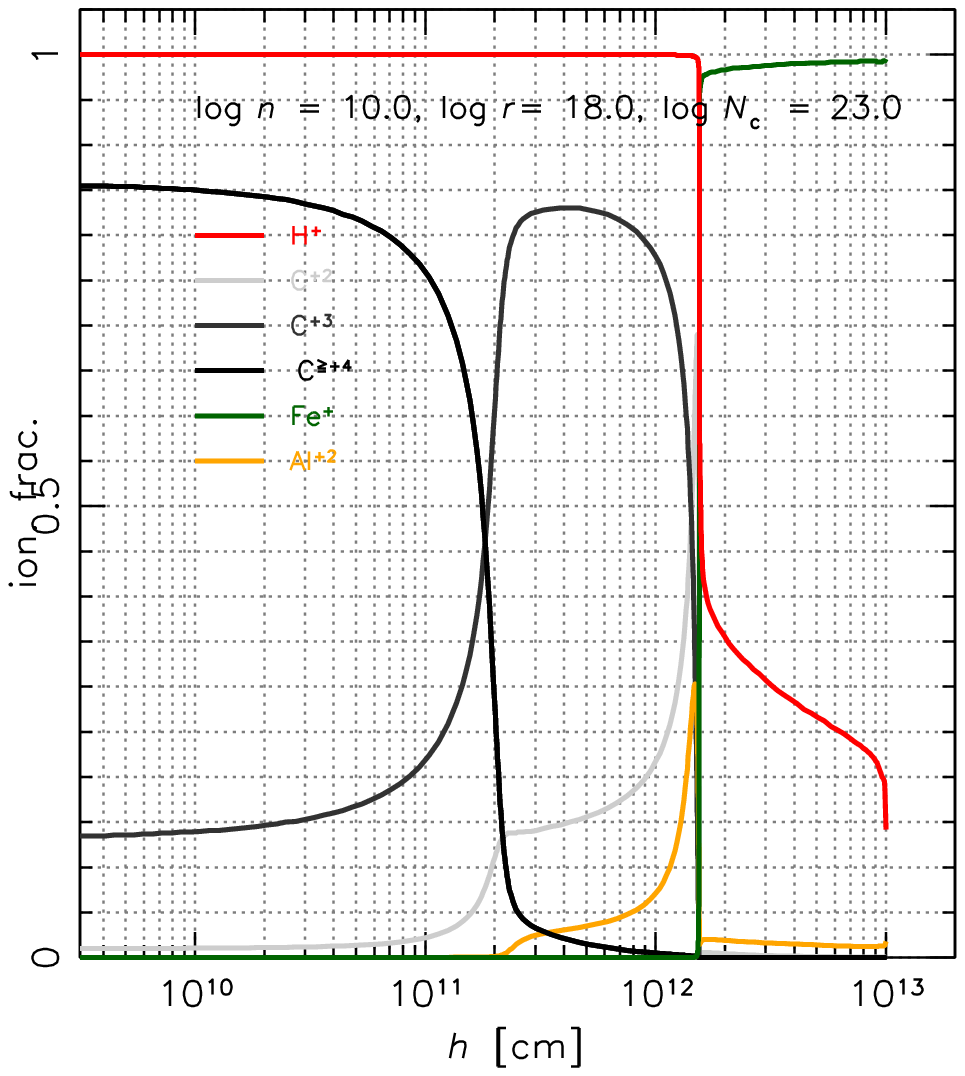}
			\includegraphics [scale = 0.30 ]{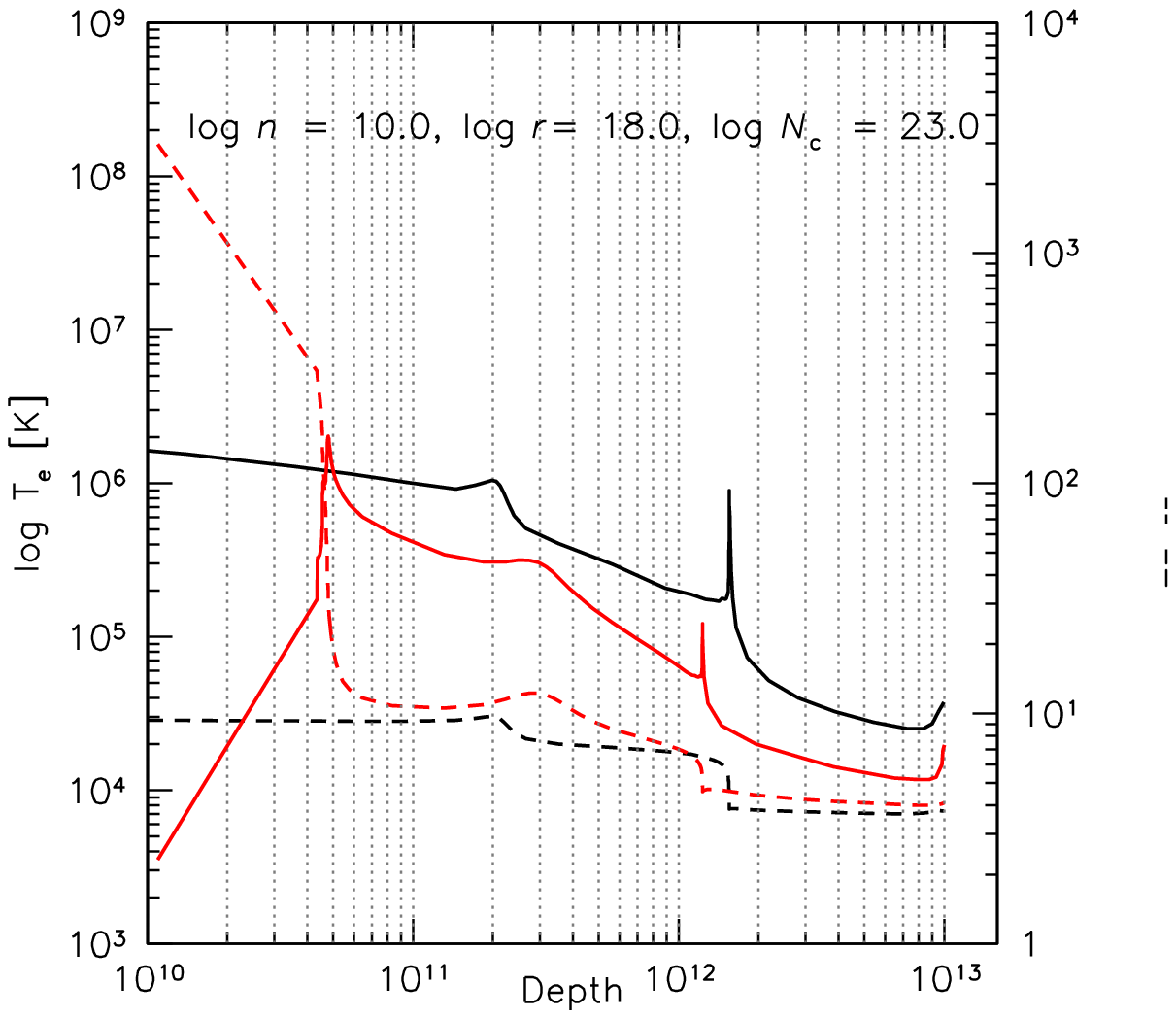}
			\caption{Ionization structure within a gas slab of column density $N_\mathrm{c} = 10^{23}$ cm$^{-2}$, log of Hydrogen density 10.0\,[cm$^{-3}$] and distance from the continuum sources $\log r = 18$ [cm] corresponding to $\approx$ 1000 $r_\mathrm{g}$, where the illuminated surface is at depth $h=0$\ (left side), for the continuum of  \object{3C 47} (left panel), and for a typical AGN continuum normalized to the same optical luminosity (middle panel). Rightmost panel: force multiplier (filled lines) and kinetic temperature (dashed lines) as a function of the geometrical depth of the gas slab for \object{3C 47} (red) and an AGN continuum (black). }
			\label{fig:ion}
	\end{figure*}

The first basic result is that the kinetic temperature is too high for a meaningful photoionization solution for moderate densities $\log n_\mathrm{H} \sim$ 9 and $\sim$ 10 at $\xi \lesssim 1000 r_\mathrm{g}$, and $\lesssim$ 500   $r_\mathrm{g}$, respectively. This implies that the bulk of the HILs might come from a region beyond the outer edge of the AD. The SED of \object{3C 47} induces an overionization of the emitting gas,  where a photoionization solution is possible. Fig. \ref{fig:ion} shows that the $C^{3+}$ ionic stage is only marginally dominant in a narrower range of depth with respect to a cloud illuminated by a typical AGN continuum, for the same optical luminosity and all other conditions kept fixed. As a consequence, the force multiplier remains systematically lower for \object{3C 47} (right panel of Fig. \ref{fig:ion}), implying that a radiation driven outflow is disfavored in this case. Another important factor hampering a radiation driven outflow is the very low \lledd: radiation forces become dominant only at low column density \citep{2010ApJ...724..318N}. Most of the optically thick gas will remain bound to the gravitational field of the black hole and reflect the Keplerian kinematics associated with the rotating disk. The modest blueward asymmetry of the \civ\ profile is consistent with a small fraction of the line emitting gas showing a radial velocity associated with outflow motions.   
		
We can constrain the radial extent of the BLR from the radius $r_{1000}$ at which the AD temperature is 1000 K to the radius of dust sublimation $r_\mathrm{dust}$ \citep{2011A&A...525L...8C}. The radius $r_{1000}$\ can be written as 
    \begin{equation}
			r_{1000} = \left(\frac{3GM\dot{M}}{8\pi \sigma_\mathrm{B} T^4} \right)^\frac{1}{3}
			\approx 6.67 \cdot 10^{16} T_{1000}^{-\frac{4}{3}} M_8^\frac{2}{3}\dot{m}_1^\frac{1}{3}\, \mathrm{cm},
	\end{equation}
		
\noindent where $\dot{M}$\ is the accretion rate,  $\sigma_\mathrm{B}$\ is the Stefan-Boltzmann constant, and $\dot{m}$ is the dimensionless accretion rate in Eddington units, i.e., $\dot{m}=\dot{M}/(L/c^2)$. For the parameters of \object{3C 47}, we obtain that $r_{1000} \approx 9.3 \times 10^{17}$\ cm $\approx 940 r_\mathrm{g}$.\ 
The ratio between $r_\mathrm{dust}$ and $r_{1000}$ is \citep[e.g.,][]{2011A&A...525L...8C}:
	\begin{equation}
			\frac{r_\mathrm{dust}}{r_{1000}} \approx 
			33 \frac{{M_8}^\frac{1}{6}}{\dot{m}_1^\frac{1}{6}\eta_{0.1}^\frac{1}{2}} \approx 12.98.
	\end{equation} 
This implies that, in the case of \object{3C 47} the BLR extension is between 1000 $r_\mathrm{g}$ and $\sim 12000 r_\mathrm{g} \approx 10^{19} $cm. At 5000 $r_\mathrm{g}$, the virial velocity is $\approx 4000$\kms, in agreement with the expected FWHM considering $\theta\approx30$. The FWHM of the symmetric Gaussian \civ\ emission feature is therefore consistent with emission concentrated toward the outer region of the permitted space. This accounts for the separation from the disk that is emitting lines from within 2000 $ r_\mathrm{g}$. 
		
Key differences between RL and RQ AGN lay in the configuration of the magnetic field and  SMBH  spin, which can significantly impact any radiation-driven wind dynamics \citep[][]{1977MNRAS.179..433B,1982MNRAS.199..883B}, though the exact roles remain debated. 
The mildly ionized wind traced by \civ\ does still exist in RLs, but is quantitatively affected \citep{1996ApJS..104...37M,2011AJ....141..167R}. The effect of the jet might be ultimately associated with a cocoon of shocked gas that develops where the pressure exerted by later expansion of the jet equals the pressure of the ambient medium. This in turn may displace the launching radius of the wind outwards \citep{2015MNRAS.450.1916S}. Therefore, we do not expect a substantial difference between RQ and RL in the occurrence of failed winds, apart from the displacement of the launching radius, and a lower force multiplier implied by a SED like the one of 3C 47, which would somewhat favor wind failure in  RLs.

\subsection{Explanations alternative to AD: a binary BLR}
\label{sec:alter}
		
An alternative scenario is the one wherein the observed double-peaked profiles could be attributed to emission from a twin BLR associated with a supermassive binary black hole. 
Under this assumption,  each of the black holes has an associated BLR, and the orbital motion of the binary system produces  Doppler shifts in the emission lines that result in the observed double-peaked profiles \citep[e.g.,][]{ 1980Natur.287..307B,1982ApJ...253..873B,1983LIACo..24..473G,1987ApJ...312L...1P,2010ApJ...709L..39C}.
The binary black hole hypothesis appears debatable on the basis of radio data. On the one hand, there is no evidence of an obvious S-shaped configuration of the jet. Actually, the 3C 47 jet extended over $\sim 200$\,kpc shows the smallest bending measures ($\lesssim$ 8 degrees) among several double-lobed sources  \citep{1994AJ....108..766B}. The position angle of the parsec scale jet mapped by VLBI \citep{2018RAA....18..108Y} is consistent with the jet mapped by the VLA. These evidence favors the stability of the jet orientation over the dynamical timescale of the radio source, $\approx 6 \cdot 10^7$\,yr \citep{2018MNRAS.474.3361T}. On the other hand, the presence of a secondary SMBH is indicated by three subtle evidence \citep{2019MNRAS.482..240K}: (1) a jet at the edge of the lobe; (2) a ring-like feature well visible in the Southern lobe \citep{1991ApJ...381...63F,1996VA.....40..173L}; (3) S-symmetry of the hot spots with respect to the jet axis, and an estimate of the geodetic precession timescale \citep{1975ApJ...199L..25B}:
    \begin{equation}
    P_\mathrm{G} \approx \frac{4 \pi c^2 }{G^\frac{3}{2}} \cdot \frac{a^\frac{5}{2} }{M^\frac{3}{2}} \cdot \frac{\tilde{q}(1+{\tilde{q}})}{(3+{4}{\tilde{q}})}\cdot(1-e^2)\sim 10^8 r_\mathrm{1}^\frac{5}{2} M_9^{-\frac{3}{2}} \mathrm{yr}
    \end{equation}
 \noindent where $\tilde{q}=M_1/M_2$\ is the mass ratio (assumed here $\tilde{q} \approx 2$), and $r = a$, the semi-major axis of the orbit for eccentricity $e = 0$, in parsecs. In the case of 3C 47, $P_\mathrm{G}$ is lower  (for \mbh$ \approx 6 \cdot 10^9$\ \msol) or comparable to the dynamic age of the radio source, and therefore compatible with the features seen in the radio lobes. 	

To examine whether this interpretation is a possible mechanism for the observed double-peaked profile in the spectra of \object{3C  47}, we recovered two additional spectra obtained in the epoch preceding 2012 (see Sect. \ref{sec:arch}). The spectra were normalized to the sum of the \oiiiseven\ NC + SBC. A previously unreported result is that the broad emission in 1996 was extremely faint and could have escaped detection if the more recent spectra had not been available for comparison: the \hb{} broad profile strength in 1996 was $\approx\frac{1}{3}$ and $\approx\frac{1}{2}$ of the ones measured on the 2006 and 2012 spectra. 
		
\begin{figure}
				\centering
				\includegraphics [scale = 0.43]{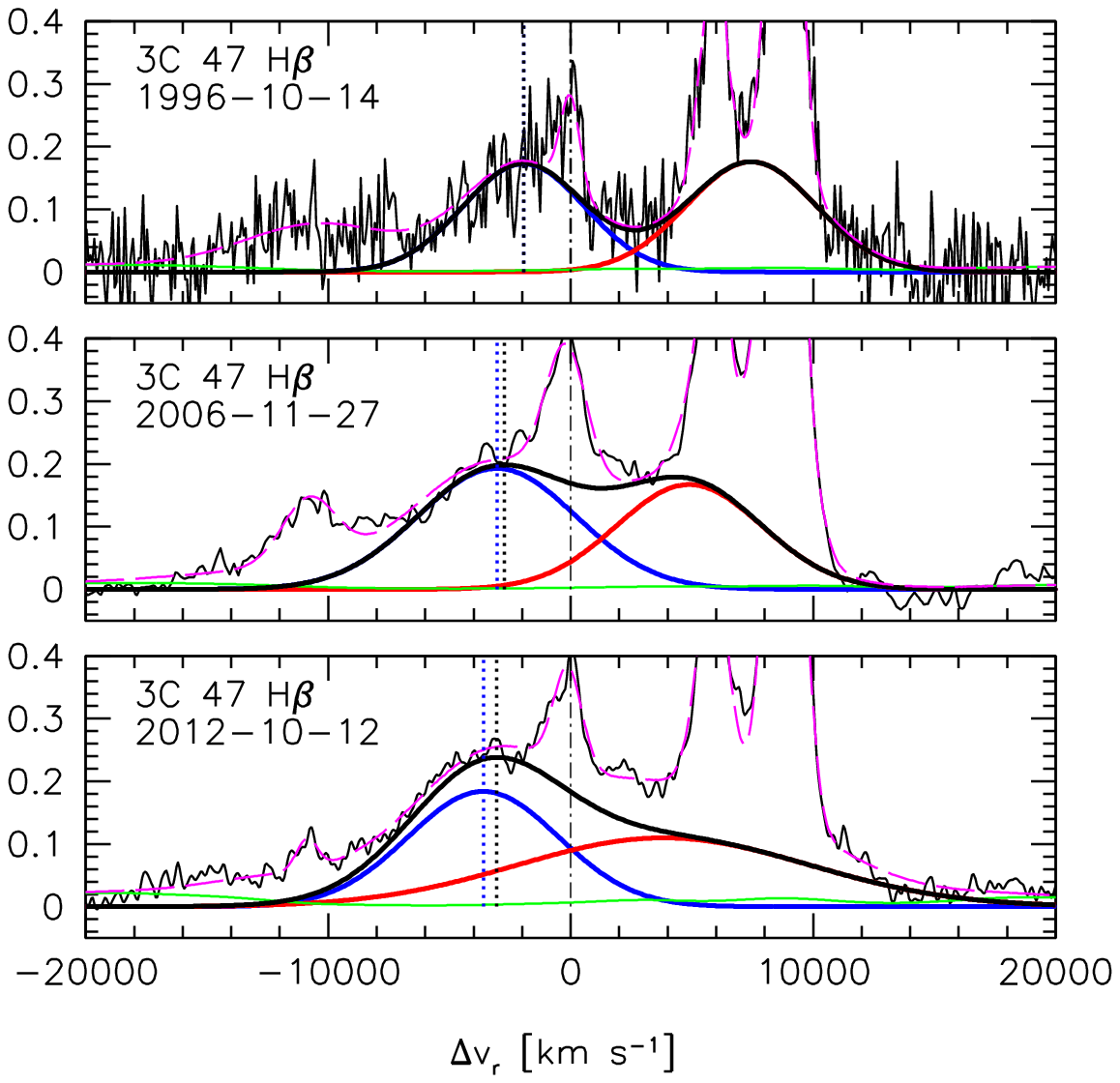}
				\includegraphics [scale = 0.43]{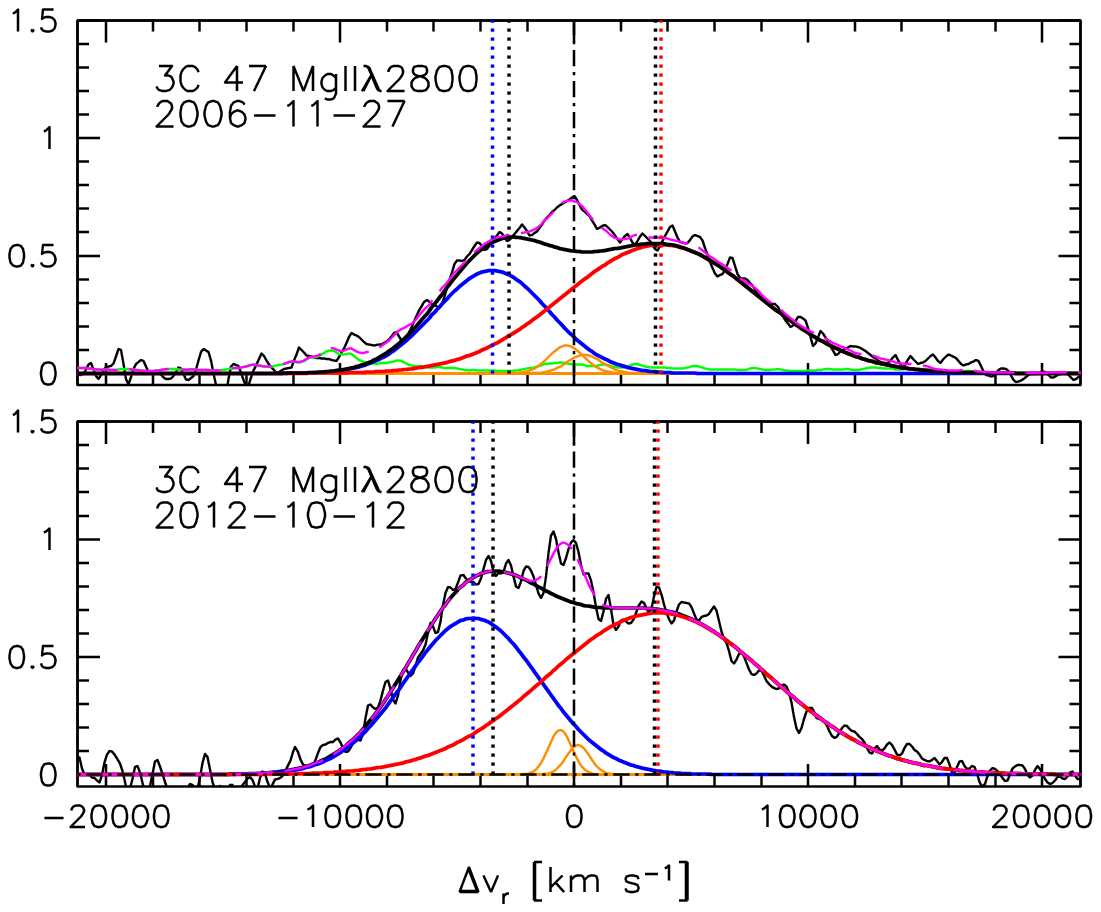}
				\vspace{-1.8cm}
				\caption{Top: Multicomponent empirical {\tt specfit} analysis results in the \hb\ line region for three epochs, after subtracting the power-law and the Balmer recombination continuum (the latter for the \mgiionly\ spectral range). The abscissa is in radial velocity units. The vertical scale corresponds to the specific flux in units of 10$^{-15}$ ergs\, s$^{-1}$\, cm$^{-2}$\, \AA$^{-1}$. The emission line components used in the fit are \feii\ (green), the blue and red peaked profiles (black), the VBC (red), and the full profile (sum of the two BCs and the VBC) (thick black), all the NCs (blue), and \oiiiopt\  SBC (orange). Black continuous lines correspond to the rest-frame spectrum. The dashed magenta line shows the model fitting from {\tt specfit}. The dot-dashed vertical lines trace the rest-frame wavelength of  \hb. Dotted lines identify the radial velocity of the blue- and redshifted broad Gaussian components. Bottom: same of the \mgiionly\ line, after removal of \feii\ emission for clarity. }
				\label {fig:emphb2}
			\end{figure}

The total mass of the binary $M = M_{1} + M_{2}$ can be estimated from the observed spectra using the third Kepler's law. We utilized the equations from \citet[][]{1997ApJ...490..216E} to estimate $M$  that correspond to the  period and orbital velocities:
	\begin{equation}
			M \sim 4.7 \times 10^{8}(1 + \tilde{q})^{3}  {P}_{100}\, v_{1,5000}^{3}\, M_{\odot},
			\label{eq:mass}
	\end{equation}
where $v_{1}$ is the orbital velocity of $M_{1}$ normalized to 5000\,\kms, and $P$ is the orbital period in units of 100\,yr. The $v_{1}\sin\theta$\ corresponds to the radial velocity of the blue peak  ($v_\mathrm{r}$ Blue) and the $v_{2}\sin \theta$\ to the red peak ($v_\mathrm{r}$ Red). We do not use $v_{2}$, since the red peak wavelength is very poorly constrained. The projected peak velocities were estimated from the empirical fittings of the lines, using two Gaussian profiles as a BC to represent the two peaks. For the \hb\ line of the 2012 spectrum (presented in this paper), the corresponding blueshift and redshift velocities were found to be $\approx$ -3200 \kms\ and $\approx$ +5000 \kms, respectively, from the peak ﬂux in the observed double-peaked profile fitted as shown in Fig. \ref{fig:emphb2} for \hb\ and \mgiionly. 
For \ha, the blue peak is located around -2800 \kms\ and the red peak is found +5900 \kms, yielding $\tilde{q}\approx 1.56$\, and $\approx 2.11$\, for \hb\ and \ha\ respectively. An average over \hb, \mgiionly\ and \ha\ at all available epochs consistently yields $\tilde{q} \approx 1.6 \pm 0.5$\ for the putative binary. The large dispersion reflects the difficulty in carrying out precise measurements of peaks of broad features affected by atmospheric absorption as well as strong overlying \oiii\ emission (red peak of \hb). The results for the measurements on \hb\ and \mgiionly\ are reported in Table \ref{tab:shifts}. The 1996 spectrum yields very poor constraints on the blueshifted peak, with $v_\mathrm{r}$\,  Blue $\approx -1900^{+500}_{-1000}$ \kms\ at 1$\sigma$ confidence level.

			\begin{table}
		\caption{Measured shifts of broad line profiles and lower limits on \mbh.} 
				\label{tab:shifts}
				\setlength{\tabcolsep}{1pt}
				\scalebox{1}{
					\begin{tabular}{@{\extracolsep{3pt}}l cccccll    @{}} 
						\hline\hline % inserting double-line
						
						Line &  \multicolumn{1}{c}{Epoch}& \multicolumn{2}{c}{Sum} && \multicolumn{2}{c}{Components} &   
						\\ \cline{3-4} \cline{6-7}
						& & \multicolumn{1}{c}{$v_\mathrm{r}$\  Blue} &$v_\mathrm{r}$\  Red 
						&& \multicolumn{1}{c}{$v_\mathrm{r}$\  Blue} &$v_\mathrm{r}$\  Red&  \\ 
						%\hline
						(1) & (2)& (3) & (4)&& (5) & (6)      \\
						\hline
						\hb\ & 1996-10-14  & -1930$^\mathrm{a}$   &   7410:  & &   -1940 &	8060: &    \\
						\hb\ & 2006-11-27  & -2730   &   4200     & &   -3040 &	4890  &             \\
						\hb\ & 2012-10-12   &-3050   &  \ldots  &  &  -3600 &	3820: &      \\ \hline
						\mgiionly\ & 2006-11-27 & -2780  &  3490   & & -3495  &	3720      &              \\
						\mgiionly\ & 2012-10-12 & -3460  &  3440   & &   -4320 &	3600 &               \\ \hline
						& $\Delta t$\ [yr] & $P$ [yr]  & \mbh\ [M$_\odot$] && P [yr]  & \mbh [M$_\odot$]\\ \hline
						\hb\  & 16.0  & 267	 & 3.8E+10$^\mathrm{b}$      && 192    &	  4.5E+10$^\mathrm{b}$\\
						\hb\  & 5.875  & 259	 & 3.7E+10$^\mathrm{b}$  && 166 &	3.9E+10$^\mathrm{b}$\\
						\mgiionly\ & 5.875 & 119 &	8.1E+09$^\mathrm{c}$ && 102 & 2.0E+10$^\mathrm{c}$\\
						\hline
					\end{tabular}
				}
				{\raggedright\\
					\textbf{Note}: $^\mathrm{a}$Highly uncertain, an F-test yields  {$v_\mathrm{r}$\  Blue $\approx -1900^{+500}_{-1000}$\ km s$^{-1}$. $^\mathrm{b}$: Assuming $(1+\tilde{q}) \approx 2.56$; $^\mathrm{c}$: Assuming $(1+\tilde{q}) \approx 2.0$. }}
			\end{table}

The only parameter that needs to be determined in  Eq. \ref{eq:mass}  is the period from the displacement of the observed spectra. Here we focus on the blueshifted peak of \hb. To estimate the period, one can consider the $\Delta v$ over the time span $\Delta t$, and the radial velocity span $v_2-v_1$\ (i.e., the separation of the two peaks). In the case of a circular orbit, the period should be $P \sim 2 (v_2 - v_1) / \Delta v \times \Delta t $,  and the values are reported for the \hb\ and \mgiionly, in both the cases of the full double-peaked profile and the fitting components. Note that $\theta$\ is known to be $\approx 20 - 30$\ degrees ( Sect. \ref{radio-opt}), so that all observed radial velocities are to be divided by $\sin \theta$\ that we assume here $\approx 30$. 

The second half of Table \ref{tab:shifts} provides the estimates of periods and total binary black hole masses from the shift variation in \hb\ and \mgiionly, over the time lapses between observations. The most constraining limit in Eq. \ref{eq:mass} comes from the measurement of the \hb\ peak: the $\Delta v \lesssim 1100$ \kms, implies a period of $P \gtrsim 200$ yr, and \mbh\ $\sim 5 \cdot 10^{10}$\ M$_\odot$,\ above the estimates of Sect. \ref{sub:accr}. The uncertainty of the 1996 measurement is large; the $-1\sigma$\ limit would imply a displacement of just 120  km s$^{-1}$\ with respect to the 2012 measurement, implying $P \sim 2800$ yr and   \mbh\ $\sim 4\times 10^{11}$ \msol. 
In addition, the measurements between 2006 and 2012 show that there is no evidence of large changes in the radial velocity of the blue peak. Even if there is an apparent systematic increase in the shift for both \hb\ and \mgii, the change in the red peak may not be consistent with orbital motion around a common center of mass. We can conclude that there is no evidence in favor of a binary BLR, given the constraints on the mass of the putative binary black hole and the uncertainties in the data. Further spectroscopic monitoring is, however, needed to fully dismiss this possibility.  

However, assuming that a second black hole is affecting the shape of the line profile in a periodic way,  the relation between the geodetic precession and orbital period can be written as: 
\begin{equation}
    P_\mathrm{G} \approx \frac{2 c^2 }{G} \cdot \frac{a}{M_1} P \cdot \tilde{q} \frac{(1+ {\tilde{q}})}{(3+ {4}{\tilde{q}})}\cdot(1-e^2) \sim 4.18\cdot 10^6 r_\mathrm{1} M_9^{-1} P_{100} \, \mathrm{yr}
    \end{equation}
The separation of the two black holes is basically unconstrained but an upper limit on $P_\mathrm{G}$\ is given by the dynamic age of the radio source, $\approx 6\cdot 10^7$\,yr. With the black hole mass of 3C 47, the age would imply a radius of $\sim 100$\,pc. Smaller radii might be possible; since 1 pc $\approx 3000$\, $r_\mathrm{g}$, it is tempting to speculate that a second black hole within a few parsecs from the primary might be responsible for the truncation of the disk at $\lesssim 2000$  $r_\mathrm{g}$.

\subsection{Further alternatives}
			
Another scenario proposed for the explanation is emission from the oppositely directed sides of a bipolar or biconical outflow. In this scenario, the double-peaked lines originate in pairs of oppositely directed cones of outflowing gas that are accelerated by the passage of the relativistic radio jets through the line emitting region and interaction with the gas immediately around the central engine \citep[e.g.,][]{1984A&A...141...85N, 1990ApJ...365..115Z,1995ApJ...438L...1S}. Evidence of line broadening after a flare-like event has been found \citep{2013ApJ...763L..36L,2020ApJ...891...68C}, and can be interpreted as an unresolved bipolar outflow. These phenomena are, however, transient over timescales of years or less. Gas moving at $\approx 4000$ \kms for slightly less than 20 years would imply a displacement $\sim 0.1$ pc. Maintaining a stable structure within the BLR in a sort of "fountain" to replenish the outflowing gas appears to be a rather unrealistic scenario. 

\subsection{3C 47 in the context of the quasar main sequence and of the RL/RQ dichotomy}

The most powerful radio sources in the local Universe belong to systems with very high \mbh\  
and low levels of activity \citep[e.g.,][]{2007ApJ...658..815S,2011MNRAS.416..917C,2017FrASS...4....1F}. 
3C 47 is not a peculiar object in this respect.  The observational and accretion properties are consistent with extreme Population B sources having very broad, double-peaked profiles, and low \feii\  emission that in turn imply large masses and low Eddington ratios. Albeit the fraction of RL sources tends to increase with FWHM in low-$z$\ samples \citep{2013ApJ...764..150M,2022MNRAS.516.2824C}, RLs remain a minority up to bin B1$^{++}$\ (12000\,\kms$\le$\,FWHM $<$\,16000, \rfeopt $< 0.5$), to which 3C 47 belongs. In this spectral bin, the fraction of irregular and multi-peaked profiles is highest along the MS \citep{2019A&A...630A.110G}.  
The usefulness of 3C 47 for understanding quasar populations along the MS and the RQ/RL dichotomy resides in the fact that, within the disk model, we can identify the radial distances over which the LILs are emitted, and, by comparing and fitting the \civ{} line, we can identify an additional emission contribution that is interpreted as due to a failed wind. This solves an open question that has been left standing since the early 1990s when HST/FOS became available for large samples of objects: why did the \civ{} broad line profile appear of comparable width or narrower than H$\beta$ in the UV spectra of radio galaxies \citep{2005MNRAS.356.1029B,2007ApJ...666..757S}? 3C 47 has offered an interpretation, thanks to the double-peaked profiles of LILs, and the large equivalent width, single-peaked CIV profile that can be accounted for by a disk profile and an almost symmetric emission with FWHM $\sim 4000$\,\kms. This phenomenon is thought to occur in both RQ and RL quasars, and may involve additional emission of Balmer lines as well \citep{2004A&A...423..909P,2009MNRAS.400..924B}. In most cases, the disk contribution could be completely masked, and empirical parameters measured on the observed profiles would appear inconsistent with the shifts and asymmetries expected from a relativistic disk \citep{1990ApJ...355L..15S}. 

\section{Conclusions}

\label{sec:concl}
In this paper, a new long-slit optical spectra of extremely jetted quasar \object{3C 47} with radio emission (log\,\rk\,>\,3) and a clearly visible double-peaked profile in low-ionization emission lines, as well as UV observations from the HST-FOS  were presented. The research work utilized an AD model, a Bayesian approach, and multicomponent non-linear fitting to analyze the resulting spectra. 
			
\indent The main findings that we draw from this study are:
			
\begin{itemize}
\item  We applied a model based on a relativistic AD  and this model successfully explained the observed double-peaked profiles of \hb, \ha, and \mgiionly.  
The agreement between the observed profile and the model is remarkable, implying that an AD is reprocessing ionizing continuum emission with 100 and 2000 gravitational radii.   The main alternative -- a double BLR associated with a binary black hole -- is found to be less appealing than the disk model for the quasar \object{3C 47}: albeit some changes in the shifts were measured, they are inconsistent with the kind of systematic variations expected for a binary model.  This is not to say that 3C 47 is necessarily a system with one SMBH. A second black hole might be present in the system, as suggested by the morphology of the radio lobes, and by a tentative estimate of the geodetic precession timescale. However, the putative presence of a second massive black hole only leads to speculative implications on the BLR structure. 
				
\item The profile of the high-ionization lines can also be modeled by the contribution of the AD, along with fairly symmetric additional components (a failed wind scenario). The failed wind scenario is supported by the lack of prominent blueshifted emission in \civ\ and in \ciii\ emission lines, although the asymmetry of the profiles still reveals a modest outflow component. Exploratory photoionization computations suggest that the gas at the radii where the AD is able to reprocess the radiation is being over-ionized for typical BLR density $n_\mathrm H \lesssim 10^{11}$ cm$^{3}$\ for the SED of \object{3C 47}, implying a lower radiative acceleration with respect to a \citet{1987ApJ...323..456M} SED for the same optical flux.  \\
				
\end{itemize}
			
The study supports not only the notion that the double-peaked profiles originate from a rotating AD  surrounding a SMBH but also the HIL  profiles -- that were not explained fully in previous studies -- are consistent with a physical scenario involving the AD and a failed outflow associated with low \lledd\ and a SED with a higher X-to-UV photon ratio with respect to the conventional RQ AGN. The presence of the symmetric component on top of the disk, associated with emission at a few thousand gravitational radii, accounts also for the difficulty in the interpretation of the \civ\ profiles, as the wind components merge smoothly with the innermost NLR profiles that are systematically broader than the \oiiiopt\ lines in most RL AGN at low-$z$ \citep{1999ApJ...518L...9S}. 
			\\
			
			%-------------------------------------- Two column figure (place early!)
			
\begin{acknowledgements}
STM acknowledges the support from Jimma University under the Ministry of Science and Higher Education. STM and MP acknowledge financial support from the Space Science and Geospatial Institute (SSGI), funded through the Ministry of Innovation and Technology (MInT). STM, ADO, and MP acknowledge financial support through the grant CSIC I-COOP+2020, COOPA20447. ADM, ADO,  JP, MP, and PM acknowledge financial support from the Spanish MCIU through projects PID2019–106027GB–C41, PID2022-140871NB-C21 by “ERDF A way of making Europe”, and the Severo Ochoa grant CEX2021- 515001131-S funded by MCIN/AEI/10.13039/501100011033. In this work, we made use of astronomical tool IRAF, which is distributed by the National Optical Astronomy Observatories, and archival data from NVSS, HST and VLA. This research is based on observations made with the 3.5m telescope at the Observatory of Calar Alto (CAHA, Almeria Spain). We thank all the CAHA staff for their high professionalism and support with the observations. Based in part on observations collected at Copernico 1.82m telescope  (Asiago Mount Ekar, Italy) INAF - Osservatorio Astronomico di Padova. This research has made use of the NASA/IPAC Extragalactic Database (NED), which is operated by the Jet Propulsion Laboratory, California Institute of Technology, under contract with the National Aeronautics and Space Administration.
				
			\end{acknowledgements}
			
			\bibliographystyle{aa} % style aa.bst
			\bibliography{1_my_references_paper_II} % your references Yourfile.bib

\end{document}